\documentclass{pas}

\usepackage{comment}
\usepackage{rotating}
\usepackage{amsmath,amsbsy}
\usepackage{graphicx}
\usepackage{rotating}
\usepackage{multirow}

\begin{document}

\lefttitle{Publications of the Astronomical Society of Australia}
\righttitle{T. W. C. Stevenson}

\jnlPage{x}{x}
\jnlDoiYr{2026}
\doival{10.1017/pasa.xxxx.xx}

\articletitt{Research Paper}

\title{Dynamic Trajectory Analysis of Meteoroids Showing Minimal Deceleration}

\author{\sn{Stevenson} \gn{T. W. C.}$^{1,2}$, \sn{Sansom} \gn{E. K.}$^{1,2}$, \sn{Devillepoix} \gn{H. A. R.}$^{1,2}$, \sn{Gritsevich} \gn{M.}$^{3,4}$, \sn{Zappatini} \gn{A.}$^{5}$, \sn{Jenniskens} \gn{P.}$^{6}$, \sn{Herd} \gn{C. D. K.}$^{7}$, \sn{Horner} \gn{J.}$^{8}$, \sn{Moskovitz} \gn{N.}$^{9}$, \sn{Hemmelgarn} \gn{S.}$^{9}$ and \sn{Daly} \gn{L.}$^{10,11,12}$}

\affil{$^1$Curtin University, Space Science and Technology Centre, 314 Wark Ave, Bentley WA 6102, Australia, $^2$International Centre for Radio Astronomy Research, 1 Turner Ave, Bentley WA 6102, Australia, $^3$Faculty of Science, University of Helsinki, Gustaf Hallströmin katu 2, FI-00014 Helsinki, Finland, $^4$Instituto de Astrofísica de Andalucia (IAA-CSIC), Glorieta de la Astronomía, Granada E-18008, Spain, $^5$Institute of Geological Sciences, University of Bern, Baltzerstrasse 3, 3012 Bern, Switzerland, $^6$SETI Institute, 339 Bernardo Ave, Mountain View, CA 94043, USA, $^7$Department of Earth and Atmospheric Sciences, 1-26 Earth Sciences Building, University of Alberta, Edmonton, Alberta, T6G 2E3, Canada, $^8$Centre for Astrophysics, University of Southern Queensland, Toowoomba, QLD 4350 Australia, $^9$Lowell Observatory, 1400 West Mars Hill Rd, Flagstaff, AZ 86001, USA, $^{10}$School of Geographical and Earth Sciences, University of Glasgow, Glasgow, G12 8QQ, UK, $^{11}$Department of Materials, University of Oxford, Oxford, OX1 3PH, UK and $^{12}$Australian Centre for Microscopy and Microanalysis, The University of Sydney, Sydney NSW 2006, Australia.}

\corresp{T. W. C. Stevenson, thomas.stevenson@curtin.edu.au}

\history{(Received xx xx xxxx; revised xx xx xxxx; accepted xx xx xxxx)}

\begin{abstract}

Meteoroids decelerate and ablate as they descend through Earth's atmosphere, however a portion of instrumentally observed meteors show little measurable deceleration and remain poorly characterised. These are referred to as `minimally decelerating objects' (MDOs). The traditional $\alpha$-$\beta$ method of dynamic trajectory analysis cannot reliably determine their preatmospheric masses or rates of ablation. We present a new approach for estimating the ballistic coefficients ($\alpha$) and mass loss parameters ($\beta$) of MDOs, allowing their inclusion in dynamic analyses. This new method employs bulk ablation coefficients (\(\sigma_b\)) derived from instrumentally observed meteorite falls and large meteor shower bodies. It is applied to MDOs comprising approximately one-third of the Global Fireball Observatory (GFO) 2014 - 2024 dataset. Our results show that MDOs are predominantly small objects occupying a distinct region of $\alpha$-$\beta$ space. Material types can be identified using supplementary data such as emission spectra, which we demonstrate using observations of 10 small iron meteoroids. This methodology expands the range of meteoroid populations accessible to dynamic trajectory analysis, providing new constraints on meteorite deposition and the compositional diversity of near-Earth objects.

\end{abstract}

\begin{keywords}
Chondrites, Fireballs, Iron meteorites, Meteor showers, Meteorites, Meteoroids, Meteors
\end{keywords}

\maketitle

\section{Introduction} \label{sec:intro}

Meteoroids ranging from \(\mu\)m to m in diameter enter the Earth system at a rate inversely proportional to their size \citep{BlandArtemieva2006sizedistribution, Silber2018meteorshockwaves, RN420}. They deliver a broad range of extraterrestrial materials, from metallic alloys to silicate minerals to cometary dust \citep{RN113, OstrowskiBryson2019meteorites}. Occasionally, metallic and stony remnants are deposited on the Earth’s surface as meteorites. They inform scientific progress because they contain valuable information regarding the chemical and thermophysical conditions present during the early Solar System \citep{RN518, Jenniskens2024solarsystempebbles}. Understanding the development of the Solar System therefore depends on understanding the materials brought to Earth from space.

Meteorites represent only the most resistant fraction of meteoroids, not the full population we aim to characterise. Most incoming meteoroids are completely vaporised during atmospheric entry and are therefore unavailable for laboratory analysis. For those that reach the ground, preservation depends strongly on mineralogy: iron meteorites are generally more resistant to weathering than stony meteorites, whereas carbonaceous chondrites are easily altered and deteriorate more rapidly \citep{LeeBland2004meteoriteweathering, RN538}. Our goal is to develop a method for classifying all extraterrestrial material entering Earth's atmosphere, including both recoverable meteorites and objects that are lost due to atmospheric vaporisation or terrestrial weathering.

\subsection{Meteoroid Data by Optical and Spectral Camera Networks} \label{subsec:alphabetaintro}

The fireballs produced by meteoroid-atmosphere interaction are best characterised when directly observed by ground-based optical camera networks. Examples include the Global Fireball Observatory (GFO; \cite{RN166}), the Global Meteor Network (GMN; \cite{RN286}), the European Fireball Network (EFN; \cite{Boro2022a}), the Spanish Meteor Network (SPMN; \cite{PenaAsensio2023spanishnetwork}), FRIPON \citep{RN552}, as well as the historic Meteorite Observation and Recovery Project (MORP; \cite{CampbellBrown2005morp}) and Prairie Network (PN; \cite{RN895}). Fireball networks carry out trajectory analyses to predict whether meteorites have been deposited, and if so, their likely landing sites \citep{Moilanen2021strewnfields, RN286, PenaAsensio2023spanishnetwork}. Between March 2014 and December 2024, 2,102 fireballs were observed by the GFO alone \citep{RN166}, of which 1,789 have been reduced with accurate entry and terminal velocities. Their progenitor meteoroids are estimated to have ranged from 1 mm to 4 m in diameter prior to ablation, including 16 events for which meteorites have been recovered. The majority (1,521) were sporadic asteroidal or cometary impactors, which are distinguished from each other based on orbital eccentricity and Tisserand's parameter (e.g. \citealt{Tancredi2014asteroidvscometorbits}). The remaining 268 have been linked to known meteor showers based on their radiants and corresponding orbits \citep{RN413}. Crucially for this study, 1,118 of the reduced fireballs showed significant deceleration, wherein the final observed velocity ($V_t$) was less than 80\% of the initial velocity ($V_e$). The remaining 671, approximately one-third of the dataset, showed minimal deceleration, which we define as a terminal velocity exceeding 80\% of the initial velocity.

At present the optical cameras employed by the GFO in Australia are supplemented by three spectral cameras from the All-Sky Meteor Orbit System (AMOS; \cite{RN892}) since 2021 \cite{devillepoix2022meteor}. Spectral systems like these enable broad classification of meteoroids as iron, stony or cometary based on their fireball emission spectra. In addition to high temporal resolution and precise instrumental calibration \citep{RN890}, this process requires a number of assumptions, which have been described by \cite{Jenniskens2007spectra} and references therein. Spectral analysis also depends upon complex calibration curves to remove wavelength-dependent atmospheric influences, such as Rayleigh scattering and aerosol concentration \citep{Jenniskens2007spectra}, as well as instrumental effects, such as vignetting and optical extinction \citep{Millman1980spectra}. Data reduction is time consuming, manually intensive, and very difficult to automate. The spatial coverage of spectral cameras is also limited by their relative rarity, thus the traditional optical systems still dominate in meteoroid tracking and identification efforts.

\subsection{Characterising Meteoroids and Minimally Decelerating Objects} \label{subsec:mdointro}

In order to automate the optical data reduction process, fireball trajectories can be characterised using the $\alpha$-$\beta$ method of dynamic trajectory analysis described by \cite{Gritsevich2007alphabeta, Gritsevich2008}. From trajectory observations, measurable physical properties can be effectively reduced to two dimensionless parameters that quantify the ability of a meteoroid to penetrate the atmosphere ($\alpha$), and its susceptibility to mass loss ($\beta$). \cite{Gritsevich2012impactconsequences} have demonstrated the utility of the ln(\(\alpha\)) vs. ln(\(\beta\)) plot, upon which it is possible to distinguish large, crater-forming impactors from smaller metallic or chondritic meteorites. It has also been proposed that clustering in $\alpha$-$\beta$ space could allow for grouping of meteoroids based on material properties. 

To date, $\alpha$ and $\beta$ cannot be determined for events where $V_t/V_e>80\%$, preventing direct comparison with other events in the $\alpha$-$\beta$ space. The aim of this study is the characterisation of meteoroids with little or no measured deceleration, for which the standard $\alpha$-$\beta$ method of \citep{RN691} is inconclusive. We will refer to these as `minimally decelerating objects' (MDOs). As stated in Section \ref{subsec:alphabetaintro}, MDOs comprise approximately one-third of all meteoroids detected by the GFO. They are significant for a comprehensive understanding of extraterrestrial materials.
 
This paper will begin by summarising the existing $\alpha$-$\beta$ method. Analytically determined $\beta$ values from significantly decelerating GFO events, including meteorite falls, are then used to calculate a suite of bulk ablation coefficients (\(\sigma_b\)). We then introduce new formulae that strongly depend upon \(\sigma_b\) values in order to calculate  $\alpha$ and $\beta$ for MDOs. Finally, we will discuss how this method can assist in understanding unrecoverable meteoroids, especially if supplementary information is available, for example from emission spectra. We will test dynamics-based meteoroid groupings in $\alpha$ - $\beta$ space for both deceleration regimes. Our study will lay a foundation for comprehensive classification of the materials delivered to Earth by meteoroids. This in turn will inform the hazards posed by near-Earth asteroids, and contribute to broader understanding of the makeup of our Solar System.

\section{Dynamic Properties of Decelerating Meteoroids} \label{sec:doalphabeta}

The $\alpha$-$\beta$ trajectory analysis method is derived from foundational meteor physics, as presented by \cite{Stulov1995aerodinamika}, then refined by \cite{Gritsevich2008}, \cite{Sansom2019alphabeta}, and others. It is used to calculate a falling meteoroid's ballistic coefficient ($\alpha$) and mass loss parameter ($\beta$) based on directly observable kinematic properties, namely altitude and velocity. Both $\alpha$ and $\beta$ are dimensionless.
These can be reliably determined for events that show a minimum of 80\% net deceleration ($V_t/V_e<80\%$) \citep{Sansom2019alphabeta, RN552}.

\subsection{Ballistic Coefficient} \label{subsec:alphadef}

The ballistic coefficient, $\alpha$, represents the ratio between the aerodynamic drag and weight forces experienced by a falling meteoroid. If $\alpha$ is high for a particular object, then that object experiences high drag relative to its weight, and is rapidly decelerated by the dense lower layers of Earth's atmosphere. If the value of $\alpha$ is low, the drag force is comparatively low, and atmospheric penetration is enhanced. Its formal definition is as follows:

\begin{equation} \label{eq:alphadef}
\alpha = \frac{c_d \rho_0 H_0 S_e}{2 M_e sin(\gamma)}
\end{equation}

Here \(c_d\) refers to the aerodynamic drag coefficient. It is typically assumed to be unity, which represents a spherical body in a hypersonic airflow regime. Next, \(\rho_0\) and \(H_0\) are respectively the critical density (1.29 kg/m\(^3\)) and critical altitude (7160 m) of an exponential atmosphere. The variables \(S_e\) and \(M_e\) are respectively the cross-sectional area and mass of the meteoroid before atmospheric entry. Area and mass can be calculated in tandem if a particular shape and density are assumed. For instance, a modified formula that uses meteoroid density to calculate initial mass has been published by \cite{Sansom2019alphabeta}, assuming a spherical shape. Finally, \(\gamma\) is the angle between the meteoroid's trajectory and the local horizon at the point of entry.

\subsection{Mass Loss Parameter} \label{subsec:betadef}

The mass loss parameter, $\beta$, is the ratio between the fraction of a meteoroid's kinetic energy that is converted into heat, and its effective enthalpy of vaporisation \citep{RN696}. A high $\beta$ value indicates a meteoroid that rapidly loses mass due to ablation and fragmentation processes. Ablation is the continuous process by which matter at the surface is melted, vaporised, ionised and stripped away by the surrounding atmosphere. Fragmentation refers to discrete breakup events that are detectable as luminous flares, as well as subtle increases in deceleration. The meteoroids that are most likely to survive their descent long enough to deposit meteorites yield low $\beta$ values and experience a degree of mass conservation. The mass loss parameter is defined by this equation:

\begin{equation} \label{eq:betadef}
\beta = \frac{(1 - \mu) \sigma_b V_e^2}{2}
\end{equation}

Here \(\mu\) is the rotational shape factor, which is typically held constant throughout a meteoroid's descent. It is typically set to \(\frac{2}{3}\), which corresponds with an object that is rotating rapidly enough for its entire surface to be evenly ablated. The bulk ablation coefficient \(\sigma_b\) indicates the amount of mass that is removed per unit of heat energy transferred (see Section \ref{subsec:sigmameth}). Finally, \(V_e\) refers to the earliest observed velocity of the meteoroid. High velocity results in higher dynamic (`ram') pressure across the leading face of the meteoroid, which accelerates mass loss \citep{RN1711}.

\subsection{Calculation of Dynamic Parameters} \label{subsec:alphabetameth}

\begin{figure*}[h]
    \centering
    \includegraphics[width=0.8\linewidth]{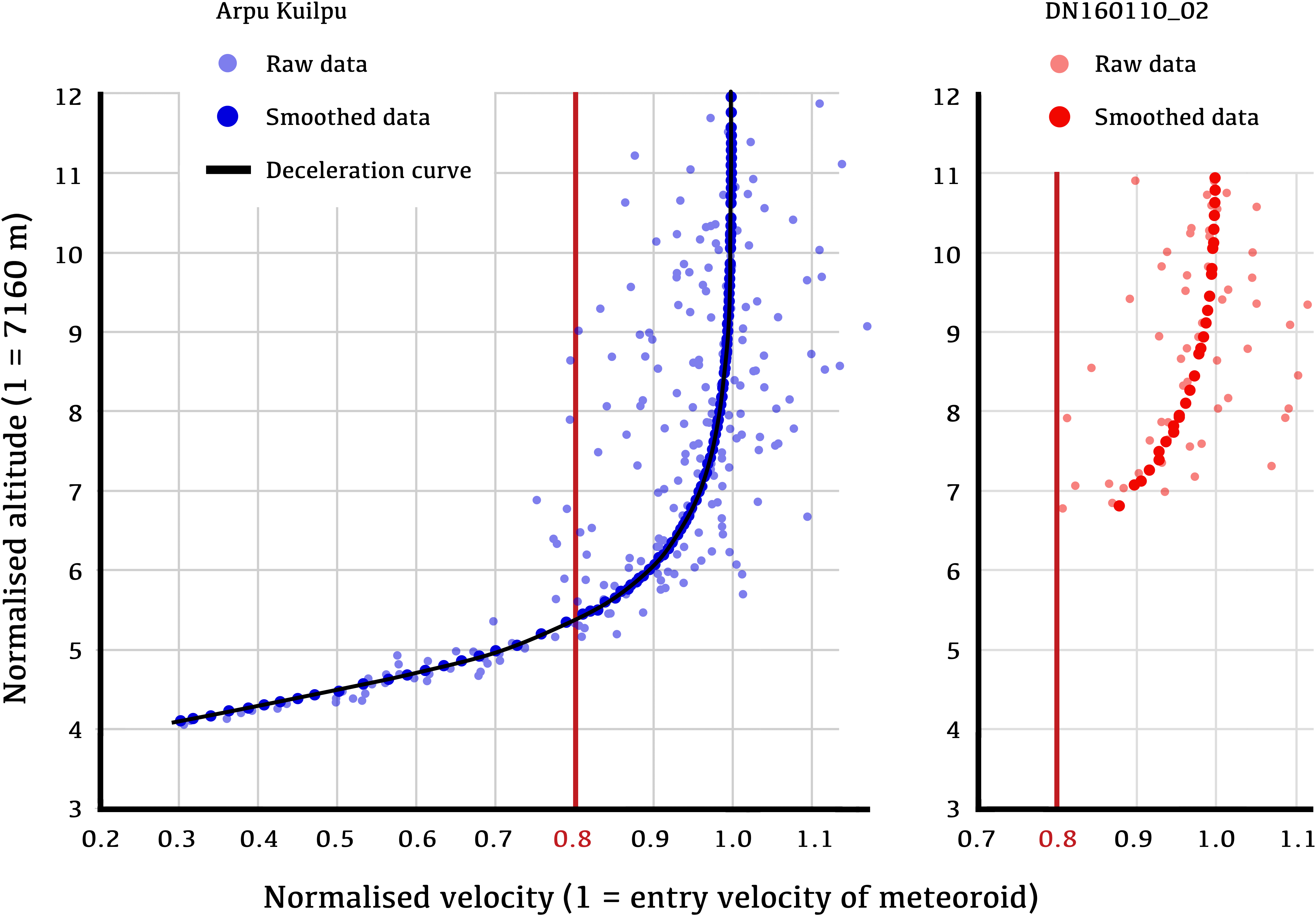}
    \caption{Normalised velocity vs. normalised altitude for the falls of Arpu Kuilpu (left) and the iron meteoroid DN240406\_02 (right). "Smoothed" data are three-point rolling averages of raw data (not to be confused with the Extended Kalman Smoother method of \cite{RN1108} used in the GFO automated pipeline).}
    \label{fig:trajectories}
\end{figure*}

For a given meteorite fall event, $\alpha$ and $\beta$ are traditionally determined analytically using normalised velocity (\(v\)) and normalised altitude (\(y\)) data (Figure \ref{fig:trajectories}). For GFO events, velocity is normalised relative to entry velocity, as determined by an Extended Kalman Smoother (EKS; \cite{RN1108}). Altitude data are normalised following the method of \cite{RN658}, which allows NRLMSISE-00 (see \cite{RN941}) or an equivalent `real' atmospheric model to be incorporated for fall site prediction purposes. We prefer a `real' atmosphere when possible; differences between our results and those of earlier literature (e.g. \cite{Sansom2019alphabeta}) can be attributed to differences in atmospheric models.

A curve is fitted to the normalised velocity vs. altitude plot using a global minimisation function. The curve includes an inflection point whose position is determined by $\alpha$. The slope of the curve below this inflection point is controlled by $\beta$. Figure \ref{fig:trajectories} includes the deceleration curve fitted to trajectory data for Arpu Kuilpu, an ordinary chondrite tracked and recovered by the GFO \citep{RN400}.

An advantage of this method is that it requires no assumptions regarding the shape, size, density or rotation of the falling meteoroid. It is only when deriving \textit{other} parameters from $\alpha$ or $\beta$, such as the preatmospheric mass or bulk ablation coefficient, that such assumptions become necessary \citep{RN691}. 
The $\alpha$-$\beta$ method provides a means of characterising objects captured by fireball networks, whether or not they survive descent.  \cite{Sansom2019alphabeta} removed the dependence of $\alpha$ upon trajectory slope to eliminate a potential bias of entry angles first identified by \cite{Gritsevich2012impactconsequences}. This is achieved by plotting \(\alpha sin \gamma\), where $\gamma$ is the entry angle measured from the local horizon. This method was applied to 273 fireballs \citep{Sansom2019alphabeta}, including three from which meteorites have been recovered, to generate a ln(\(\alpha sin \gamma\)) vs. ln(\(\beta\)) plot and determine probabilistic thresholds for meteorite deposition (`survivability curves'). Both studies agree that meteorite falls tend to yield low ballistic coefficients and low mass loss parameters when compared with the majority of fireball-producing meteoroids.

\subsection{Ablation Coefficient} \label{subsec:sigmameth}

Once $\alpha$ and $\beta$ are known, their defining equations can be rearranged and solved to determine linked parameters, given reasonable assumptions of shape, size, density and rotation. For our purposes, the most pertinent material parameter is the bulk ablation coefficient, \(\sigma_b\). It can be estimated by rearranging the equation for $\beta$, then substituting a rotational shape factor (\(\mu\)) and earliest observed velocity (\(V_e\)):

\begin{equation} \label{eq:sigmab}
\sigma_b = \frac{2 \beta}{(1 - \mu) V_e^2}
\end{equation}

An exact value for $\mu$ can be derived from light curve analysis, as proposed by \cite{GritsevichKoschny2011luminousefficiency}. When a sufficiently detailed light curve is not available, we assume \(\mu = \frac{2}{3}\) to indicate uniform ablation across the entire surface.

Ablation coefficients are sometimes considered diagnostic of particular meteoroid types \citep{RN365}, and are expressed in either kg MJ\(^{-1}\) or s\(^2\) km\(^{-2}\); these units are equivalent. We will provide \(\sigma_b\) units as kg MJ\(^{-1}\).

\subsubsection{Intrinsic versus Bulk Ablation Coefficients} \label{subsubsec:twosigmas}

An \textit{intrinsic} ablation coefficient (\(\sigma_i\)) is treated as an inherent property specific to each meteoroid type. It can be calculated using the thermophysical properties of meteoritic materials such as silicates, organics and iron-nickel alloys. Such properties include the enthalpy of vaporisation, heat transfer coefficient, and atmospheric drag coefficient \citep{PecinaCeplecha1983}. In real meteoroid entries, these values vary with temperature and are difficult or impossible to determine directly, thus they are usually assumed. For instance, the enthalpy of vaporisation for iron meteoroids is typically assumed to be equal to that of pure iron, which is a necessary simplification \citep{ReVelleCeplecha1994}. It is also possible to estimate \(\sigma_i\) values by modelling ablation and fragmentation processes separately, for example in the popular erosion simulator of \cite{Boro2007}. In terms of time and manual effort, this process is highly intensive \citep{Buccongello2024}, but has been shown to produce reliable results for well-studied meteorite falls (e.g: Winchcombe, \cite{RN1711}).

By contrast, a \textit{bulk} or \textit{apparent} ablation coefficient (\(\sigma_b\)) is calculated using direct fireball observations, and is influenced by multiple mass loss processes, namely ablation, sudden fragmentation, and `quasi-continuous' fragmentation \citep{Boro2007}. The term `ablation coefficient' is therefore not strictly accurate because it is influenced by multiple mass loss processes; hence the term `apparent'. It is typically assumed to be constant throughout a meteoroid's descent, and is unique to that particular fall event, with a minimum limit that depends upon bulk composition \citep{PecinaCeplecha1983}. It tends to be inflated by up to one order of magnitude compared to \(\sigma_i\) for events in which extensive or catastrophic fragmentation occurred (e.g: Chelyabinsk, \cite{Trigo2021}). Some models (e.g: \citealt{RN1674}) have treated \(\sigma_i\) as equivalent to \(\sigma_b\) by assuming the absence of fragmentation. However, only rare monolithic meteorites (e.g: Carancas; \cite{BoroSpurny2008}) are thought to have impacted without experiencing fragmentation within the atmosphere. It is therefore unrealistic to assume that \(\sigma_i \approx  \sigma_b\); these are rarely interchangeable. 

The $\alpha$-$\beta$ method of \cite{Gritsevich2008} yields a form of \(\sigma_b\) because it fits a deceleration curve to raw data, which includes the deceleration associated with fragmentation during luminous flight. In this paper, unless otherwise specified, we will present only bulk ablation coefficients derived from dynamic trajectory analysis. These are most appropriate to employ in our study of MDOs, which directly builds upon the works of \cite{Gritsevich2008} and \cite{RN696}.

\subsubsection{Calculated Bulk Ablation Coefficients for Chondrite Falls} \label{subsubsec:sigmachondrite}

\begin{figure*}[h]
    \centering
    \includegraphics[width=1\linewidth]{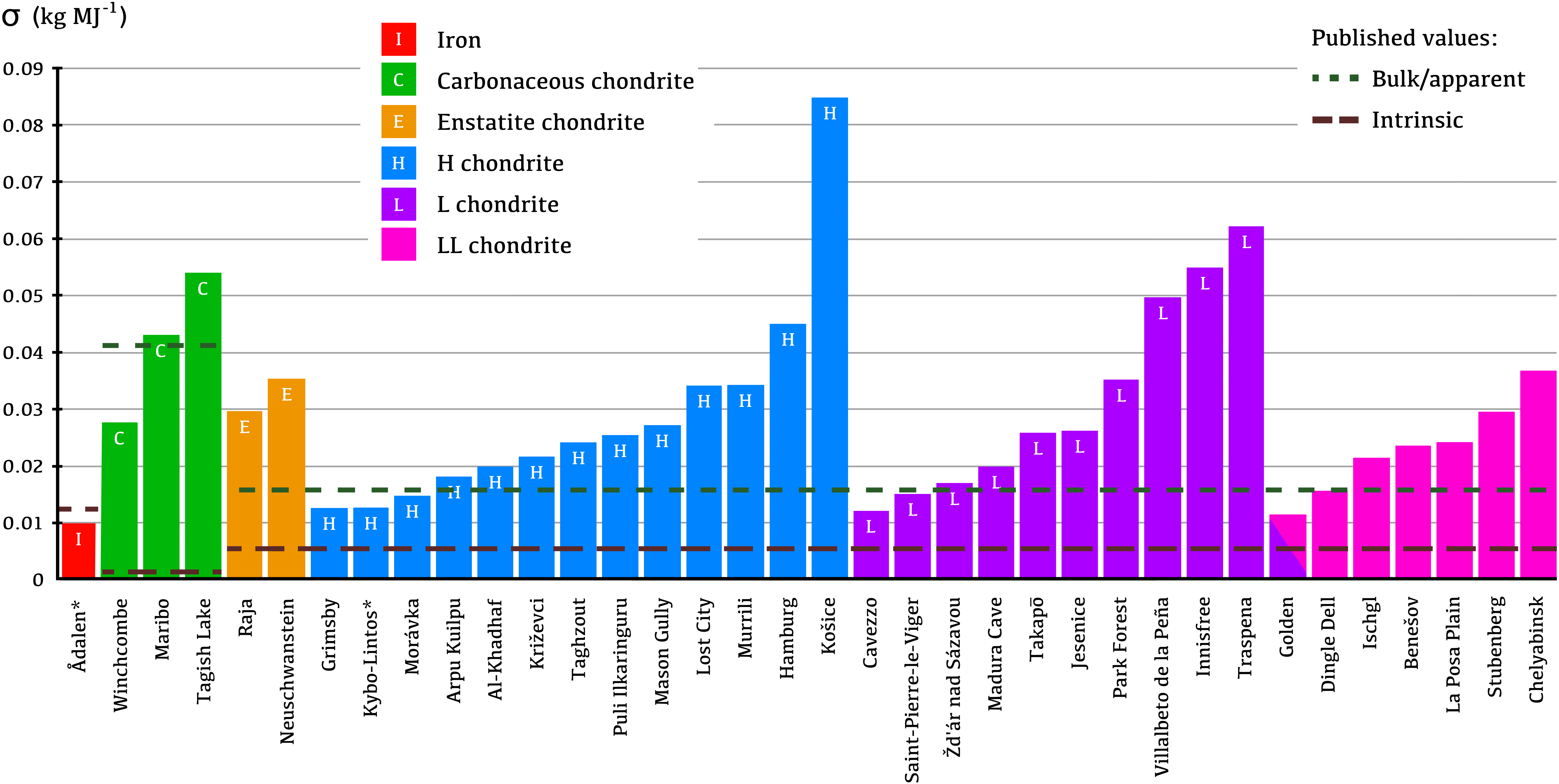}
    \caption{Bulk ablation coefficients calculated for a selection of instrumentally observed meteorite falls, assuming uniform ablation (\(\mu = \frac{2}{3}\)). * indicates a provisional name. Previously published values are shown for stony and carbonaceous \(\sigma_b\) from \citet{Ceplecha1994meteoroidproperties}, for iron \(\sigma_i\) from \citet{ReVelleCeplecha1994}, for stony \(\sigma_i\) from \citet{CeplechaReVelle2005fragmentationmodel}  and for carbonaceous \(\sigma_i\) from \citet{Ceplecha2007tagishlakefragmentation}.}
    \label{fig:sigmahisto}
\end{figure*}

In addition to GFO data, we have independently calculated mass loss parameters and bulk ablation coefficients for a global dataset of 36 meteorites with resolved atmospheric trajectories (Appendix \ref{sec:data}) using the traditional analytical fit method. Of the fallen meteorites, 35 are chondritic. As shown in Figure \ref{fig:sigmahisto}, \(\sigma_b\) is highly variable among all chondrite types. The H-chondrite Košice \citep{Kuznetsova2014kosicetrajectory} provides a notable outlier. It is believed that this meteoroid produced an airburst and experienced extensive fragmentation throughout its descent, losing comparatively little mass via pure ablation \citep{RN131, Moilanen2021strewnfields}. The majority of chondrites tend to break up in two distinct stages as they encounter dynamic pressure thresholds on the order of 0.1 and 1 MPa \citep{RN95}.

Of the three ordinary chondrite classes, H-chondrites tend to experience the lowest degree of mass loss, which is reflected by their mean \(\sigma_b\) of 0.0236 kg MJ\(^{-1}\) (s.d. 0.0099 kg MJ\(^{-1}\)), excluding Košice. Based on averages, LL-chondrites (mean 0.0262 kg MJ\(^{-1}\), s.d. 0.0063 kg MJ\(^{-1}\)) and L-chondrites (mean 0.0289 kg MJ\(^{-1}\), s.d. 0.0133 kg MJ\(^{-1}\)) are progressively less resistant to mass loss when compared with H-chondrites. Among all three ordinary chondrite groups, median \(\sigma_b\) is taken as 0.0250 kg MJ\(^{-1}\), which will be employed in our analysis of minimally decelerating asteroidal bodies (Section \ref{sec:mdometh}).

The friable carbonaceous chondrites Maribo and Tagish Lake produce \(\sigma_b\) values of 0.0430 and 0.0540 kg MJ\(^{-1}\) respectively. They fall within the upper range of possible values for ordinary chondrites. Winchcombe provides the lowest carbonaceous value known to date, 0.0276 kg MJ\(^{-1}\). This meteorite experienced relatively low dynamic pressures due to its slow entry velocity, which reduced the rate of mass loss and spared it from total destruction \citep{RN1711}.

The enstatite chondrites Raja (EH3) and Neuschwanstein (EL6) yield 0.0296 and 0.0353 kg MJ\(^{-1}\) respectively. Neuschwanstein experienced a singular fragmentation event at the lower end of its trajectory, producing several large fragments of high mechanical strength \citep{Oberst2004}. The slightly lower value obtained for Raja implies a less destructive fragmentation regime and/or a difference in ablation behaviour due to high metal content in comparison with Neuschwanstein.

Particularly astute readers may note that our \(\sigma_b\) value for Neuschwanstein is exactly three times that calculated by \cite{Gritsevich2008}. This is because, whilst we have assumed uniform ablation across all surfaces due to meteoroid rotation, the earlier study assumed no rotation (\(\mu\) = 0) and produced \(\sigma_b\) = 0.0117 kg MJ\(^{-1}\) for the same entry velocity. In conjunction, these two approaches yield the full range of possible bulk ablation coefficients for a given fall. The values illustrated in Figure \ref{fig:sigmahisto} are maxima; the corresponding minima are one-third in magnitude.

Our \(\sigma_b\) calculations for chondrites are higher than the \(\sigma_i\) values presented by \cite{RN357} for equivalent meteoroid types, but generally agree with more recent models and plasma wind tunnel experiments (e.g: \cite{RN873}). It is clear that \(\sigma_b\) can vary significantly among meteorites belonging to the same compositional group (Figure \ref{fig:sigmahisto}). We attribute this variation to differences in fragmentation and ablation behaviour primarily caused by differing sizes, shapes, entry angles and atmospheric conditions. Consequently, it is difficult to select a single bulk ablation coefficient to represent all meteoroids of the same type for the purposes of dynamic trajectory analysis. Accurate characterisation of fallen meteorites requires that each fall be considered independently of the others.

\subsubsection{Ablation Coefficients for Iron Meteoroids} \label{subsubsec:sigmairon}

Ablation coefficients for iron and stony-iron meteoroids are poorly constrained because metal-rich meteorite falls are rarely observed. Furthermore, the mechanisms by which they shed mass are not fully understood \citep{RN455}. It has been proposed that iron meteoroids experience two possible ablation regimes, one dominated by melting and the other by vaporisation, depending on their mass. \cite{ReVelleCeplecha1994} propose \(\sigma_i\) values of 0.0743 kg MJ\(^{-1}\) for iron bodies below \(2 \times 10^5\) kg, for which melting is the dominant ablation process. For larger bodies, in which vaporisation dominates, a smaller \(\sigma_i\) value of 0.0124 kg MJ\(^{-1}\) is proposed. Both of these values are approximate; they were derived from thermophysical properties of pure iron. More recently, \cite{RN99} employed the erosion model of \cite{Boro2007} to calculate \(\sigma_i\) values for four instrumentally observed iron meteoroids identified using emission spectra. One produced a value of 0.080 kg MJ\(^{-1}\), the other three 0.010 - 0.013 kg MJ\(^{-1}\). These values generally agree with the dichotomy presented by \cite{ReVelleCeplecha1994}. However, none of the four irons were expected to exceed a few mm in diameter, so the value of 0.080 kg MJ\(^{-1}\) is the only example that aligns with the prior study.

To date a single iron meteorite, provisionally named `Ådalen', has been captured by fireball cameras in sufficient detail to reconstruct its trajectory and calculate $\alpha$ and $\beta$ \citep{RN543}. Its derived bulk ablation coefficient of 0.010 kg MJ\(^{-1}\) matches the lower end of the \(\sigma_i\) range derived by \cite{RN99}. It is also marginally lower than the lower intrinsic value proposed by \cite{ReVelleCeplecha1994}: 0.0124 kg MJ\(^{-1}\). The same study states that this ablation coefficient applies only to iron meteoroids exceeding \(2 \times 10^5\) kg in entry mass, whilst the entry mass of Ådalen has been estimated at only 3500 - 8500 kg \citep{RN543, McFadden2024fireballacoustics}. This discrepancy suggests a lower actual mass boundary between melting-dominated and vaporisation-dominated ablation, or errors in the underlying theoretical calculations. In the absence of additional reliable bulk values, we will employ the \(\sigma_i\) values published by \cite{ReVelleCeplecha1994} as reasonable approximations of \(\sigma_b\) for both large and small irons.

\subsubsection{Ablation Coefficients for Meteor Shower Bodies} \label{subsubsec:sigmashower}

Ablation coefficients for meteor shower bodies (MSBs) have historically been difficult to determine due to material fragility and frequent lack of visible deceleration. Several MSBs captured by the GFO do show significant deceleration, allowing $\alpha$ and $\beta$ to be determined using the traditional method. \(\sigma_b\) has been derived for these objects using Equation \ref{eq:sigmab} and assuming uniform ablation (\(\mu=\frac{2}{3}\)), which is the most common situation according to \cite{Bouquet2014orbitaldetectionsimulation}.

Table 1 displays mean \(\sigma_b\) values for significantly decelerating objects associated with meteor showers for which multiple such objects have been detected. MSB \(\sigma_i\) values are available in \cite{Buccongello2024}; we recommend consulting their paper for a thorough review of meteor shower properties and provenance.

\begin{table*}[h]
    \begin{center}
    \caption{Ablation coefficients of significantly decelerating objects associated with selected meteor showers. Bulk coefficients (\(\sigma_b\)) calculated from GFO detections between 2014 - 2024. Intrinsic coefficients (\(\sigma_i\)) sourced from \cite{Buccongello2024}. s.d.: standard deviation; \textit{n}: number of objects.}
    \begin{tabular}{|l|l|l|l|}
    \hline
     Meteor shower & Likely parent & \(\sigma_b\) (kg MJ\(^{-1}\)) & \(\sigma_i\) (kg MJ\(^{-1}\)) \\
     \hline
     Geminids & 3200 Phaethon & 0.020; s.d. 0.011 & 0.028; s.d. 0.002 \\
     (GEM) & \citep{RN865} & \textit{n=15} & \textit{n=5} \\
     \null & \citep{RN1607} & \null & \null \\
    \hline
     \(\alpha\)--Capricornids & 169P/NEAT & 0.026; s.d. 0.014 & 0.017; s.d. 0.004 \\
     (CAP) & \citep{RN1612} & \textit{n=3} & \textit{n=3} \\
    \hline
     \(\eta\)--Virginids & 2003 FB5 & 0.031; s.d. 0.018 & \null \\
     (EVI) & \citep{Jenniskens2023meteorshoweratlas} & \textit{n=2} & \null \\
     \null & \citep{Borovicka2026etavirginids} & \null & \null \\
    \hline
     October \(\alpha\)--Virginids & 1998 SH2 & 0.015; s.d. 0.001 & \null \\
     (OAV) & \citep{RN1615} & \textit{n=2} & \null \\
    \hline
     Northern Taurids & 2004 TG10 & 0.049; s.d. 0.028 & 0.027; s.d. 0.002 \\
     (NTA) & \citep{Jenniskens2023meteorshoweratlas} & \textit{n=5} & \textit{n=2} \\
    \hline
     Southern Taurids & 2P/Encke & 0.050; s.d. 0.027 & 0.027; s.d. 0.002 \\
     (STA) & \citep{Porubcan2009} & \textit{n=6} & \textit{n=7} \\
    \hline
     Other showers & \null & 0.043; s.d. 0.029 & \null \\
     \null & \null & \textit{n=12} & \null \\
     \hline
    \end{tabular}
    \label{tab:sigmashower}
    \end{center}
\end{table*}

For the Geminids, the mean value of \(\sigma_b\) is lower than that of \(\sigma_i\), contrary to the pattern observed for stony meteorite falls and objects associated with the other meteor showers in Table 1. Objects originating from the Geminids shower are believed to approximate carbonaceous chondrites in bulk composition \citep{RN26, RN865}. There are numerous theorised differences between these and meteorite progenitors that could contribute to the apparent reduction of \(\sigma_b\) compared to \(\sigma_i\), such as smaller sizes and densities, higher porosities, and faster entry velocities. A deeper discussion of MSB structure lies beyond the scope of this study.

\subsection{Results of Decelerating Meteoroid Types} \label{subsec:docomp}

\begin{figure*}[h]
    \centering
    \includegraphics[width=1\linewidth]{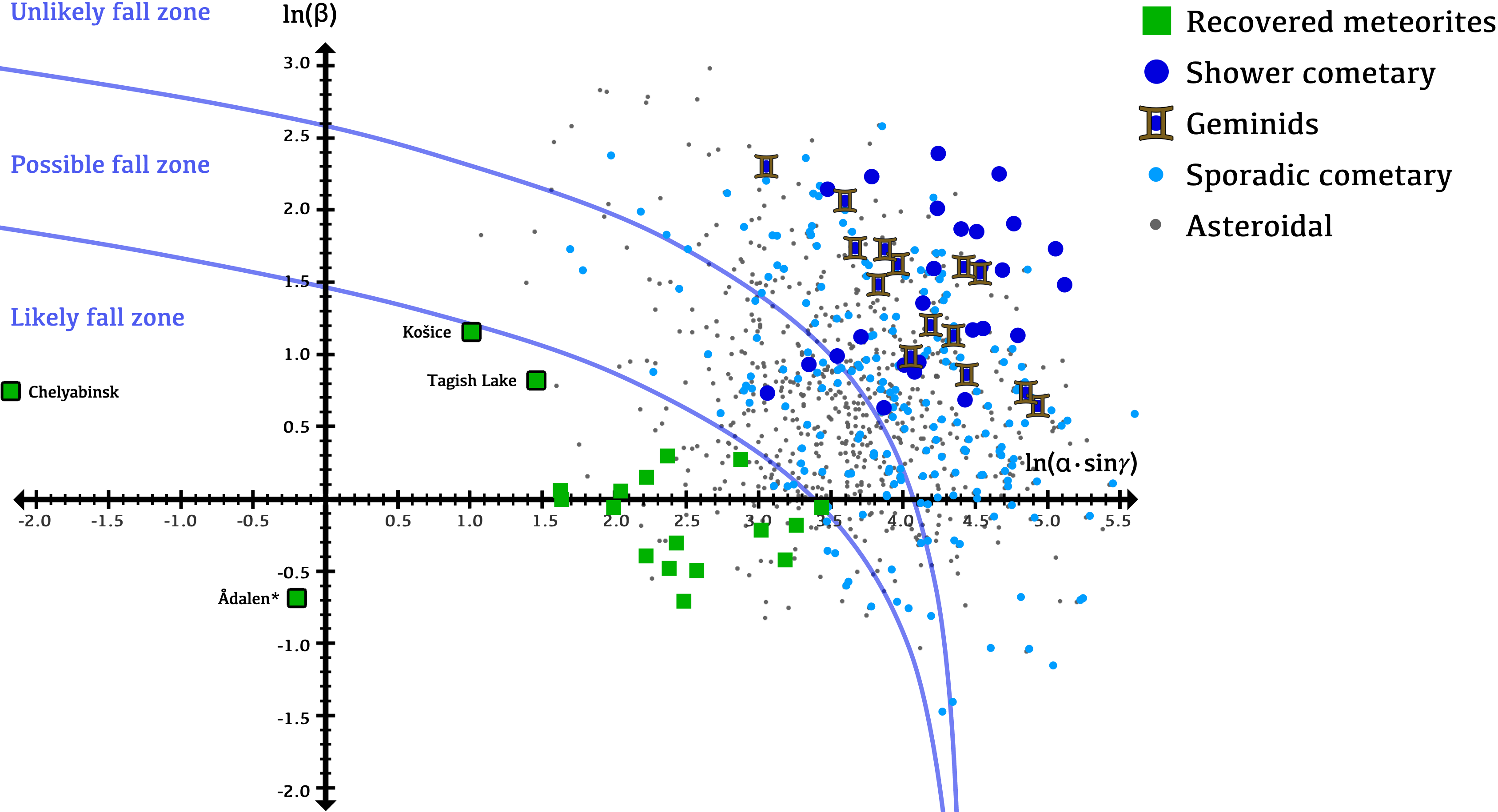}
    \caption{Ballistic coefficient (with slope dependence removed, \(\ln(\alpha sin\gamma)\)) vs. logarithmic mass loss parameter (\(\ln(\beta)\)) for significantly decelerating meteoroids captured by the GFO between March 2014 and December 2024. Non-GFO meteorites Chelyabinsk, Košice, Tagish Lake and Ådalen* are included for comparison. Survivability curves of \cite{Sansom2019alphabeta} are given for 50 g meteorites. * indicates a provisional name. Note that `asteroidal' vs. `cometary' classifications are defined by orbital parameters only.}
    \label{fig:doalphabeta}
\end{figure*}

The reduced GFO 2014 - 2024 dataset contains 1,118 events that showed sufficient deceleration for the standard $\alpha$-$\beta$ method to be applied. Of these, 45 are associated with meteor showers and another 16 with fall trajectories of recovered meteorites from participating nations (\href{https://dfn.gfo.rocks/meteorites.html}{https://dfn.gfo.rocks/meteorites.html}). The remainder have been analysed and classified as either asteroidal or cometary on the basis of orbital properties \citep{RN413}. They include many suspected meteorite droppers from which solid fragments have not yet been recovered. Since all of these objects experienced a significant degree of deceleration, their $\alpha$ and $\beta$ values have been calculated using the method described in Section \ref{subsec:alphabetameth}. They are displayed in Figure \ref{fig:doalphabeta}, an update of Figure 3 from \cite{Sansom2019alphabeta}. We have adopted the convention proposed by the same study to remove the influence of entry angle by plotting \(\ln(\alpha sin\gamma)\) instead of \(\ln(\alpha)\). This removes potential bias against objects with a comparatively shallow entry angle and consequently long flight time. We also include the survivability curves devised by \cite{Sansom2019alphabeta} for a terminal mass of 50 g or more. Figure \ref{fig:doalphabeta} lends credibility to these survivability curves, showing recovered meteorites are indeed only found in the `Likely' and `Possible' fall zones.

\section{Methods for Determining $\alpha$ and $\beta$ for Minimally Decelerating Objects} \label{sec:mdometh}

\subsection{Minimally Decelerating Objects in the Global Fireball Observatory Dataset} \label{subsec:gfomdo}

The standard $\alpha$-$\beta$ method relies upon an analytical fit to normalised trajectory data. A reasonable degree of deceleration must be present for the fit to converge upon a reliable solution, such that the final observed velocity of an object is less than 80\% of its entry velocity. If this is not the case, we consider it a `minimally decelerating object' (MDO) upon which the standard method is unreliable. Figure \ref{fig:trajectories} illustrates the practicality of fitting a curve to a trajectory characterised by `significant' versus `minimal' deceleration, even when smoothed data are employed. After a meteoroid has slowed to below 80\% of its entry velocity, as in the case of Arpu Kuilpu, the inflection point in its trajectory is fully visible and the slope of the lower portion can be tightly constrained. By contrast, if a meteoroid does not reach the 80\% threshold, then the inflection point cannot be fully resolved and the lower portion of the trajectory cannot be constrained with any degree of accuracy. This is illustrated by the iron meteoroid DN240406\_02 (Figure \ref{fig:trajectories}).

Over one-third of all fireballs tracked by the GFO can be considered MDOs. Of the 671 MDO falls, 418 represent sporadic impactors, and 253 are associated with known meteor showers. We expect the majority of MDOs to be small, sub-cm bodies that are completely vaporised at relatively high altitudes. Total vaporisation is typical of fast-moving, poorly consolidated cometary remnants. Since atmospheric density is low at high altitude, these objects are destroyed before the atmosphere can exert significant deceleration upon them. Terminal altitudes (\(H_t\)) and velocities (\(V_t\)) are always high, and terminal masses (\(M_t\)) are always zero.

\subsection{Novel Formulae for the Ballistic Coefficient and Mass Loss Parameter} \label{subsec:neweqs}

To better characterise MDOs, the following logarithmic equation from \citet{RN696} is employed to estimate $\alpha$ and $\beta$, wherein the variable \(y_t\) represents the normalised terminal altitude. This formula was derived from fundamental meteor physics equations under the assumption of zero terminal mass (\(M_t = 0\)). It effectively states that the terminal altitude of a totally vaporised meteoroid is proportional to the product of its ballistic coefficient and mass loss parameter:

\begin{equation} \label{eq:gritspope}
y_t = \ln(2 \alpha \beta)
\end{equation}

\bigskip

\begin{figure*}[h]
    \centering
    \includegraphics[width=0.7\linewidth]{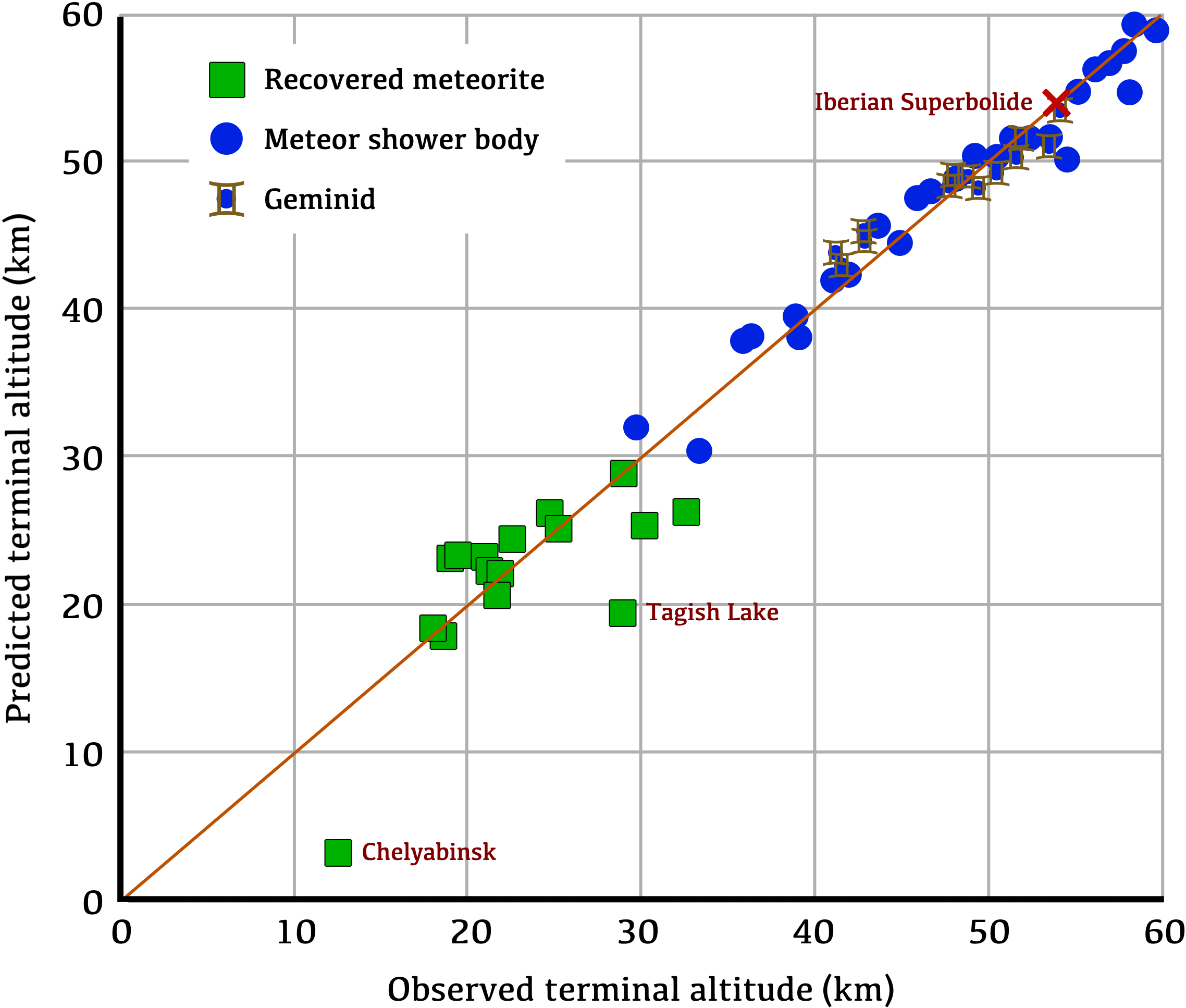}
    \caption{Observed vs. predicted terminal altitudes for GFO-detected meteoroids of known composition showing significant deceleration. Predicted terminal altitudes are calculated using \(H_0 ln (2 \alpha \beta)\), where \(H_0\) is the standard altitude normalisation factor of 7.16 km. Chelyabinsk, Tagish Lake and the 2024 Iberian Superbolide are included for comparison.}
    \label{fig:plotvalid}
\end{figure*}

\cite{RN664} validated this formula by applying it to small meteoroids detected by MORP \citep{Halliday1996morp, CampbellBrown2005morp}. We have applied the same formula to meteoroids within the GFO dataset and found the same correlation displayed in Figure 1 of \cite{RN664}. Interestingly, this statement holds true for many past events in which a non-zero terminal mass was deposited. Figure~\ref{fig:plotvalid} compares observed terminal altitudes with those predicted by Equation \ref{eq:gritspope} for GFO decelerating bodies with known compositions and independently calculated $\alpha$ and $\beta$ values. Both the recovered meteorites and vaporised MSBs fall within a 10\% error margin of the 1:1 equivalency line. Note the position of the 2024 Iberian Superbolide \citep{10.1093/mnrasl/slae065}, which falls exactly on the equivalency line and is believed to have been totally reduced to dust.

As demonstrated by Figure \ref{fig:plotvalid}, Equation \ref{eq:gritspope} is appropriate for relatively small meteoroids. It does not hold for objects with large preatmospheric masses, including the airburst-producing meteoroids Chelyabinsk \citep{Trigo2021} and Tagish Lake \citep{Ceplecha2007tagishlakefragmentation}. We are confident that it provides reasonable $\alpha$ and $\beta$ estimates for MDOs, which tend to be much smaller than meteorite droppers and airburst generators.

Further derivation (see Appendix \ref{sec:deriv} for full details) produces the following formulae for $\alpha$ and $\beta$, given the additional assumption of uniform ablation across the entire meteoroid surface:

\begin{equation} \label{eq:mdoalpha}
\alpha = \frac{3 e^{y_t}}{\sigma_b V_e^2}
\end{equation}

\begin{equation} \label{eq:mdobeta}
\beta = \frac{e^{y_t}}{2 \alpha} = \frac{\sigma_b V_e^2}{6}
\end{equation}

Note that the value of $\beta$ depends upon the value of $\alpha$, which in turn depends upon \(\sigma_b\), the bulk ablation coefficient. For any given value of \(\sigma_b\), both $\alpha$ and $\beta$ can be derived under the assumption of zero terminal mass. The better \(\sigma_b\) values are constrained, the more accurate our dynamic characterisation of MDOs becomes. 253 of our MDOs are associated with meteor showers, thus we have used the \(\sigma_b\) values recorded in Table 1 to estimate their dynamic parameters. The 418 sporadic MDOs have been classified as asteroidal or cometary on the basis of their calculated preatmospheric orbital elements, namely Tisserand's parameter and eccentricity \citep{Tancredi2014asteroidvscometorbits}. Asteroidal MDOs have been assigned a bulk ablation coefficient of 0.025 kg MJ\(^{-1}\); the median value for chondritic meteorites. For sporadic cometary MDOs, \(\sigma_b\) = 0.050 kg MJ\(^{-1}\). This is the mean bulk value obtained from a subset of 6 significantly decelerating objects with highly elliptical orbits but no shower association.

We focus on a set of 10 MDOs detected over Australia and continental Europe that have been assigned iron compositions, based upon spectra analysed by the AMOS and EFN teams \citep{RN1304, RN892}. The spectra have been matched with accurate fall trajectories provided by the GFO, or by \cite{RN1304, RN99, Boro2022b} for European cases. Entry velocities and terminal altitudes are well-constrained in all 10 cases. Each of these iron meteoroids has an estimated entry mass of 150 g or less, and no surviving terminal mass. We use Equations \ref{eq:mdoalpha} and \ref{eq:mdobeta} to determine minimally decelerating $\alpha$ and $\beta$ for these objects, and in accordance with the work of \cite{ReVelleCeplecha1994} and references therein, we assume a melting-dominated ablation coefficient of 0.0743 kg MJ\(^{-1}\).

\subsection{Results of Minimally Decelerating Objects} \label{subsec:mdodisc}

\begin{figure*}[h]
    \centering
    \includegraphics[width=1\linewidth]{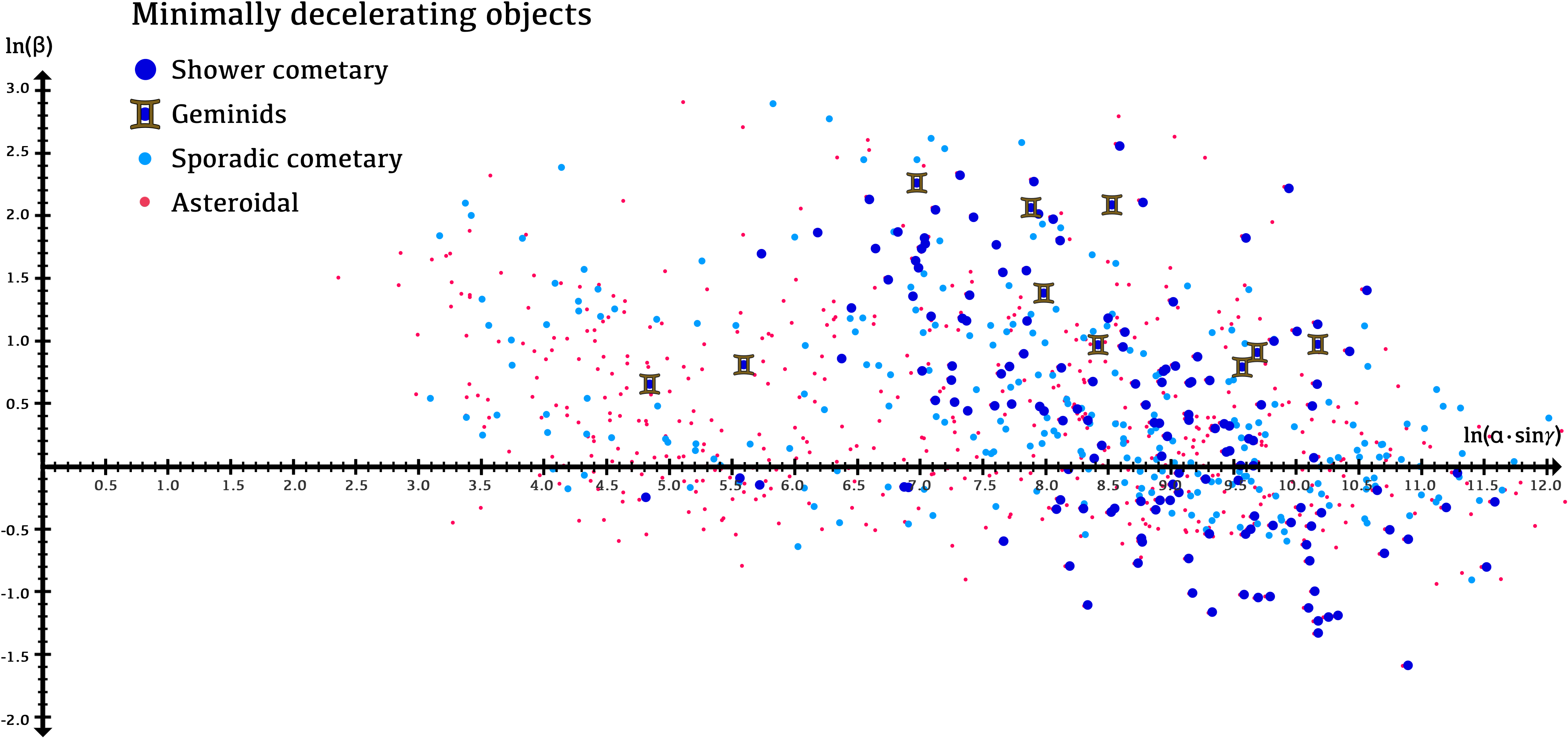}
    \caption{\(\ln(\alpha sin\gamma)\) vs. \(\ln(\beta)\) for minimally decelerating meteoroids captured by the GFO between March 2014 and December 2024.}
    \label{fig:mdoalphabeta}
\end{figure*}

Figure \ref{fig:mdoalphabeta} shows the positions of all GFO-detected MDOs in $\alpha$-$\beta$ space. The position of each object relies on an estimated bulk ablation coefficient, $\sigma_b$, as explained in the previous section. This visualisation can be used as a tool for understanding meteoroid dynamic properties and the likelihood of meteorite deposition from a given fall event.

\begin{figure*}[h]
    \centering
    \includegraphics[width=1\linewidth]{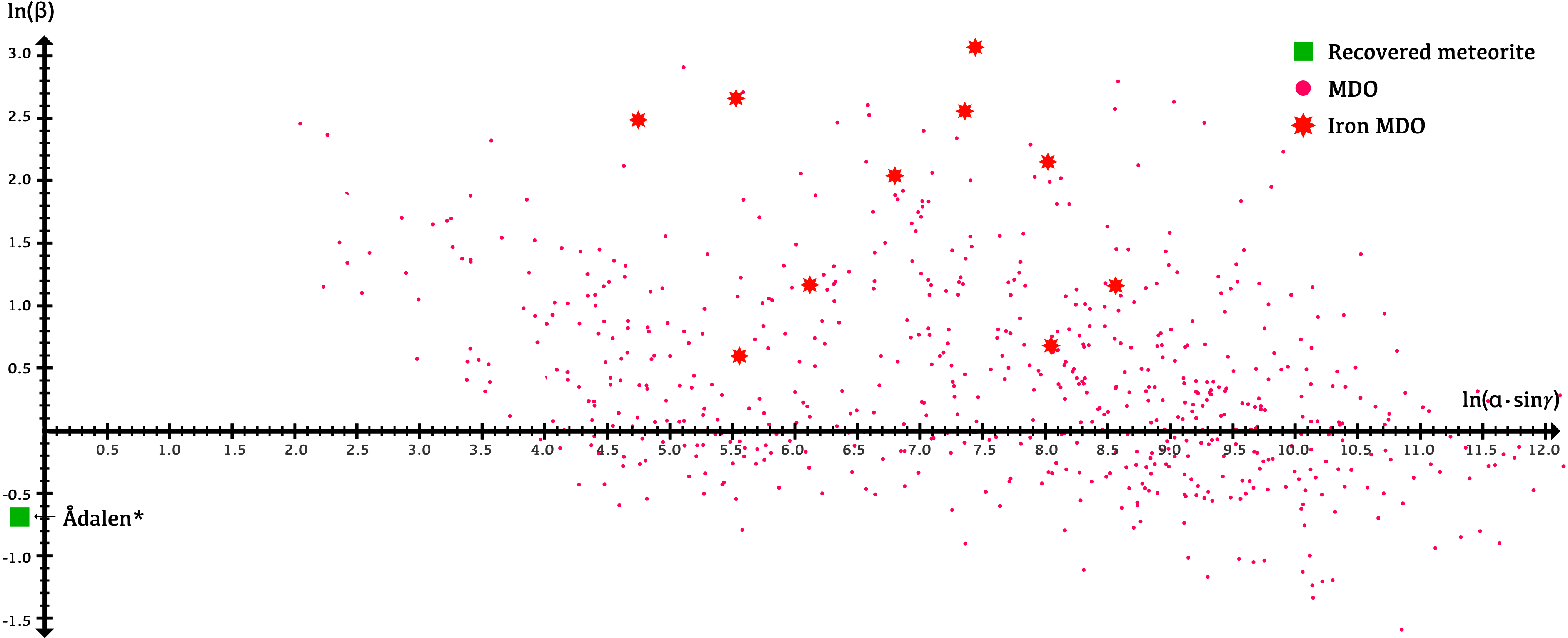}
    \caption{Figure \ref{fig:mdoalphabeta} with minimally decelerating iron meteoroids highlighted. The iron meteorite Ådalen is included for comparison.}
    \label{fig:irons}
\end{figure*}

Initially, this study hoped to identify spatial variation among these data that could be attributed to variation in material properties, namely density. To investigate this hypothesis, MDOs of known composition can be used to test whether or not meteoroids' distribution in $\alpha$-$\beta$ space depends upon their density. For this purpose, detailed emission spectra are invaluable for determining bulk compositions, as in the case of the 10 AMOS/EFN iron meteoroids highlighted in Figure \ref{fig:irons}.

\section{Discussion}

\subsection{Bulk vs. Intrinsic Ablation Parameters}

In this study it has been important to distinguish between bulk (\(\sigma_b\)) and intrinsic (\(\sigma_i\)) ablation coefficients, particularly given previous studies have published bulk or intrinsic values without necessarily differentiating them. The intrinsic ablation coefficient (\(\sigma_i\)) is  an inherent property of a given meteoroid, based on thermophysical characteristics. For instrumentally observed meteoroid entries, the ablation parameter is more commonly determined as an apparent (or `bulk') value. This bulk ablation parameter (\(\sigma_b\)) incorporates not only ablation, but also other mass loss effects, such as fragmentation. 
From our calculations of \(\sigma_b\) for chondrite falls, small iron meteoroids and meteor shower bodies, there is an interesting relation that emerges between \(\sigma_i\) and \(\sigma_b\). When ablation is the dominant mass loss mode, we expect similar values of \(\sigma_i\) and \(\sigma_b\); this is the case for iron meteoroids. Some chondritic falls also yield $\sigma_i \approx \sigma_b$, although most show fragmentation as the dominant mass loss mode and yield $\sigma_i < \sigma_b$. 

Meteor shower bodies originating from cometary streams also show $\sigma_i < \sigma_b$. No cometary remnants are confirmed to have survived the passage through Earth's atmosphere. In spite of this fact, enough is understood about such MSBs to explain their measured dynamic parameters. Comets are primarily composed of ices of water, methane, ammonia and carbon dioxide, mixed with organic and/or siliceous particles \citep{RN467, Wesolowski2025cometanalog}. The volatility of the icy components promotes their vaporisation and removal during solar approach \citep{RN1224}, so that cometary remnants become progressively more porous and volatile-depleted over successive orbits. We expect this tenuous structure to promote a relatively high rate of mass loss via fragmentation, rather than pure ablation, during atmospheric entry.

\subsection{Density/Mass Sorting of Significantly Decelerating Objects}\label{sec:disc_do}

Among significantly decelerating objects (Figure \ref{fig:doalphabeta}), the size characteristics of a given meteoroid can generally be correlated with its ballistic coefficient. This is primarily due to the influences of bulk density and preatmospheric mass. The higher the density or mass, the lower the corresponding value of $\alpha$. Objects of greater density contain more mass per unit of cross-sectional area. Consequently, they experience lower drag in comparison to their weight and penetrate deeper into the atmosphere. Meteoroids with a relatively dense, stony/asteroidal composition are therefore more common in the low-$\alpha$ region than bodies of cometary origin.

The highly friable nature of cometary remnants, coupled with the high heliocentric velocities imparted by their parent comets, can explain the high $\beta$ values obtained for MSBs in comparison with stony objects. Furthermore, compressed air exerting pressure upon the void spaces inside a porous cometary body magnifies aerodynamic drag, producing a high $\alpha$ value and further enhancing fragmentation \citep{RN1371}. 

Figure \ref{fig:doalphabeta} shows that larger and denser objects tend to plot toward lower $\alpha$ and $\beta$ values. Significantly decelerating objects associated with the Geminid meteoroid stream do not align with the overall density trend described above. They are thought to be denser than cometary remnants \citep{Buccongello2024, Jenniskens2024meteorshoweratlas}, yet they do not produce lower $\alpha$ values as expected of stony bodies, instead showing similar characteristics to cometary remnants. Geminids are fragments of the asteroid 3200 Phaethon \citep{RN1607}, and are expected to be similar to carbonaceous chondrites in terms of bulk density and composition \citep{RN26}. However, Figure \ref{fig:doalphabeta} demonstrates their high ballistic coefficients compared with other stony objects, and even with many cometary MSBs, which are extremely porous and light. It is possible that Geminids are also highly porous, to a degree not considered in any dynamic trajectory analysis. Alternatively, they may represent minor icy components of their predominantly rocky parent asteroid, although this scenario is unlikely due to the asteroid's close proximity to the Sun at perihelion, and the Geminids' small sizes precluding volatile preservation \citep{RN865, RN862}. The JAXA-operated DESTINY+ mission, which will visit 3200 Phaethon and its associated meteoroid stream in 2028 \citep{2018LPI....49.2570A}, may shed light on this mystery.

\subsection{Sorting of Minimally Decelerating Objects}

There are 671 minimally decelerating objects (MDOs) among the reduced GFO dataset, with 253 determined to be from meteor showers, and 10 known from spectral analyses to be iron meteoroids. These are smaller objects for which a mass/density sorting or clustering effect may not be visible, contrary to the significantly decelerating bodies discussed in Section \ref{sec:disc_do}.

It is clear from Figure \ref{fig:irons} that `minimally decelerating' iron meteoroids do not occupy the same region of $\alpha$-$\beta$ space as the much larger object Ådalen. They yield much higher ballistic coefficients because of their small size and high drag-to-weight ratio. Their mass loss parameters are higher because they have been assigned a relatively high ablation coefficient, in accordance with \cite{ReVelleCeplecha1994}. Consequently, these meteoroids blend into the region occupied by MDOs of stony or cometary composition. There is no apparent trend in their positions. They have been identified as iron meteoroids on the basis of their emission spectra, using the methods and underlying assumptions described by \cite{Jenniskens2007spectra} and \cite{RN890}. In the absence of spectra, this level of compositional/density distinction would not have been possible. Herein lies a compelling reason for the continued development and expansion of spectral systems within global fireball networks. Our results also illustrate that the ability of our MDO formulae to distinguish between different compositions and densities is not straightforward, unless data from alternative sources or analyses are available.

We observe that MDOs occupy a wider range of possible ballistic coefficients than other meteoroids do. Their wide spread along the horizontal axis is likely due to a size sorting effect, rather than density variation. The majority of MDOs are minuscule bodies with expected entry masses well below 1 kg, which generate high aerodynamic drag forces in comparison to their weight forces. Although there is some overlap between these and decelerating objects, the majority of MDOs occupy positions beyond \(\ln(\alpha \sin(\gamma)) = 5.5\) where no kg-scale meteoroids have yet been recorded.

Where overlap does occur, it is in the region where \(2.8 < \ln(\alpha \sin(\gamma)) < 5.5\). Most of the bodies in this region are sporadic cometary or asteroidal remnants. There is only one object (DN141214\_08) with a confirmed meteor shower association in this region, which belongs to the Geminids. We have examined this object in detail to elucidate factors that separate it from other bodies originating from the Geminids shower. Note that the closest observing camera was situated over 240 km from the fireball, thus any conclusions based on this observation should be considered with healthy scepticism.

Although the traditional $\alpha$ - $\beta$ method tends to fail for MDOs, certain dynamic parameters such as preatmospheric mass can be estimated using the Extended Kalman Smoother (EKS) built into the GFO reduction pipeline. For the Geminid, an initial mass of 4 kg has been derived. Compared with other Geminids detected by the same network, this mass is relatively large. We take 4 kg to be the upper limit possible for an MDO; several confirmed meteorites including Arpu Kuilpu \citep{RN371} and Cavezzo \citep{Gardiol2021cavezzo} yield entry mass estimates below this threshold. There is nothing else visibly unusual about the DN141214\_08; its entry and terminal altitudes and velocities are typical for the Geminids shower. If this object was indeed unusually large, it may also have been unusually weak, undergoing a singular catastrophic fragmentation event at high altitude.

To explain the overlap between normal and minimally decelerating populations in $\alpha$-$\beta$ space, one must first consider instrumentation and analytical error. It is also possible that our $\sigma_b$ estimates are too inaccurate. Otherwise, the overlap can be explained by objects of unusually high entry mass (hence low $\alpha$) coupled with low mechanical strength, such as `rubble piles' and `dust balls'. These objects could deposit a large number of fragments into the atmosphere, but the small sizes of individual fragments would promote total vaporisation before significant deceleration could occur.

\section{Conclusions} \label{sec:conc}

Dynamic trajectory analysis using the $\alpha$-$\beta$ method, as initially presented by \cite{Gritsevich2008}, provides a baseline for classification of Earth-impacting meteoroids, whether or not they survive impact. It provides a physically grounded system to classify a wide range of meteoroids, including those that deposit meteorites on the ground. In previous studies this method was limited to meteoroids showing significant deceleration during atmospheric descent. 1,118 such meteoroids were detected by the GFO between March 2014 and December 2024. For the first time, we have estimated $\alpha$ and $\beta$ values for `minimally decelerating' objects (MDOs) using a method based upon the terminal altitude formula of \cite{RN696} and reasonable bulk ablation coefficient estimates ($\sigma_b$). The GFO detected 671 MDOs in the 2014 - 2024 period of operation, of which 253 have been associated with known meteor showers. The parameters thus derived can provide insights into the shape, strength, ablation efficiency, and mineralogical nature of shower and sporadic meteoroids. 

The most important findings of this study are as follows:

\begin{enumerate}

\item Ballistic coefficients ($\alpha$) and mass loss parameters ($\beta$) can be estimated for MDOs under the assumption of zero terminal mass, using a reasonable estimate of the bulk/apparent ablation coefficient (\(\sigma_b\)).

\item Bulk ablation coefficients for a range of meteoroid compositions have been constrained using $\beta$ values from a suite of decelerating meteoroids, including 36 instrumentally observed meteorite falls, and 15 events associated with meteor shower bodies. They reflect fragmentation processes in addition to pure ablation. For stony meteorites, \(\sigma_b\) can be highly variable and tends to exceed the intrinsic ablation coefficient \(\sigma_i\) (Figure \ref{fig:sigmahisto}). For meteor shower bodies, \(\sigma_b\) tends to be higher than \(\sigma_i\), with the exception of the Geminids (Table 1).

\item Observed and theoretical ablation coefficients for iron meteoroids have been directly compared. Previously, intrinsic ablation coefficients of 0.0743 and 0.0124 kg MJ\(^{-1}\) were proposed for irons respectively smaller or larger than $2.0 \times 10^5$ kg. Our derived bulk value of 0.010 kg MJ\(^{-1}\) for Ådalen challenges this dichotomy because its estimated entry mass ranges from $3.5 - 8.5 \times 10^3$ kg. Further study of iron meteorites in the laboratory is required to more accurately constrain ablation behaviours, in particular the threshold size between melting- and vaporisation-dominated regimes.

\item There is an inverse correlation between the bulk density of a significantly decelerating object and its $\alpha$ value, the primary exceptions being large airburst producers and Geminids (Figure \ref{fig:doalphabeta}). This relationship is not valid for MDOs.

\item The position of an object in $\alpha$-$\beta$ space can be used to evaluate its likelihood of depositing meteorites, but not its bulk composition. This finding reinforces the `fate criterion' of \cite{Sansom2019alphabeta}.

\end{enumerate}

To achieve dynamics-based classification of all extraterrestrial impactors, there is a continual demand for linked kinematic, dynamic and compositional data from meteorite falls. In particular, observations of iron meteorites are lacking due to the apparent rarity of large iron meteoroids in near-Earth space. Planned expansion of spectral camera systems across Australia will enable the addition of elemental data to our analyses. 

The ability to resolve spectra over time, for example by linking them with traditional fireball cameras, will also allow for easy identification of fragmentation points. The timing of fragmentation can then be used as a proxy for bulk mechanical strength. In a future study we will combine $\alpha$, $\beta$, composition and an appropriate strength indicator to classify large numbers of detected meteoroids. This line of research will enable a more comprehensive understanding of the objects raining upon Earth from space.

\section{Acknowledgements}

The primary authors acknowledge the traditional custodians of the land where they operate, the Whadjuk Nyoongar people of Western Australia, and their elders past, present and future.

The Desert Fireball Network and Global Fireball Observatory programs have been funded by the Australian Research Council as part of the Discovery Project scheme (DP170102529, DP200102073, DP230100301), the Linkage Infrastructure, Equipment and Facilities scheme (LE170100106), and institutional support from Curtin University. 

We thank Drs P. G. Brown and J. Borovička for their valuable feedback during the preparation of this paper.

T. W. C. Stevenson acknowledges the support of an Australian Government Research Training Program (RTP) Scholarship to support his PhD candidacy.

M. Gritsevich expresses gratitude to the Academy of Finland for supporting the project no. 325806 (PlanetS), which facilitated the development of the analytical methods presented in this paper. The program of development within Priority-2030 is acknowledged for supporting researches at UrFU.

L. Daly acknowledges support from UKRI STFC grants (ST/Y004817/1, ST/T002328/1, ST/W001128/1, and ST/V000799/1).

\bibliographystyle{apalike}
\bibliography{MDOpaper}

@ARTICLE{ReVelleCeplecha1994,
       author = {ReVelle, D. O. and Ceplecha, Z.},
        title = "{Analysis of identified iron meteoroids: {P}ossible relation with {M}-type {E}arth-crossing asteroids?}",
      journal = {Astronomy and Astrophysics},
         year = 1994,
        month = dec,
       volume = {292},
        pages = {330-336},
       adsurl = {https://ui.adsabs.harvard.edu/abs/1994AandA...292..330R}
}

@article{Oberst2004,
author = {Oberst, J. and Heinlein, D. and Köhler, U. and Spurný, P.},
title = {The multiple meteorite fall of {N}euschwanstein: {C}ircumstances of the event and meteorite search campaigns},
journal = {Meteoritics and Planetary Science},
volume = {39},
number = {10},
pages = {1627-1641},
doi = {https://doi.org/10.1111/j.1945-5100.2004.tb00062.x},
url = {https://onlinelibrary.wiley.com/doi/abs/10.1111/j.1945-5100.2004.tb00062.x},
eprint = {https://onlinelibrary.wiley.com/doi/pdf/10.1111/j.1945-5100.2004.tb00062.x},
year = {2004}
}

@article{BoroSpurny2008,
	author = {Borovička, J. and Spurný, P.},
	title = {The {C}arancas meteorite impact: {E}ncounter  with a monolithic meteoroid},
	doi= {10.1051/0004-6361:200809905},
	url= {https://doi.org/10.1051/0004-6361:200809905},
	journal = {Astronomy and Astrophysics},
	year = 2008,
	volume = 485,
	number = 2,
	pages = "L1-L4",
}

@article{Trigo2021,
author = {Trigo-Rodríguez, Josep M. and Dergham, Joan and Gritsevich, Maria and Lyytinen, Esko and Silber, Elizabeth A. and Williams, Iwan P.},
title = {A Numerical Approach to Study Ablation of Large Bolides: {A}pplication to {C}helyabinsk},
journal = {Advances in Astronomy},
volume = {2021},
number = {1},
pages = {8852772},
doi = {https://doi.org/10.1155/2021/8852772},
url = {https://onlinelibrary.wiley.com/doi/abs/10.1155/2021/8852772},
eprint = {https://onlinelibrary.wiley.com/doi/pdf/10.1155/2021/8852772},
year = {2021}
}

@INPROCEEDINGS{2018LPI....49.2570A,
       author = {{Arai}, T. and {Kobayashi}, M. and {Ishibashi}, K. and {Yoshida}, F. and {Kimura}, H. and {Wada}, K. and {Senshu}, H. and {Yamada}, M. and {Okudaira}, O. and {Okamoto}, T. and {Kameda}, S. and {Srama}, R. and {Kruger}, H. and {Ishiguro}, M. and {Yabuta}, H. and {Nakamura}, T. and {Watanabe}, J. and {Ito}, T. and {Ohtsuka}, K. and {Tachibana}, S. and {Mikouchi}, T. and {Komatsu}, M. and {Nakamura-Messenger}, K. and {Sasaki}, S. and {Hiroi}, T. and {Abe}, S. and {Urakawa}, S. and {Hirata}, N. and {Demura}, H. and {Komatsu}, G. and {Noguchi}, T. and {Sekiguchi}, T. and {Inamori}, T. and {Yano}, H. and {Yoshikawa}, M. and {Ohtsubo}, T. and {Okada}, T. and {Iwata}, T. and {Nishiyama}, K. and {Toyota}, T. and {Kawakatsu}, Y. and {Takashima}, T.},
        title = "{{DESTINY}+ Mission: {F}lyby of {G}eminids Parent Asteroid (3200) {P}haethon and In-Situ Analyses of Dust Accreting on the {E}arth}",
    booktitle = {49th Annual Lunar and Planetary Science Conference},
         year = 2018,
       series = {Lunar and Planetary Science Conference},
        month = mar,
          eid = {2570},
        pages = {2570},
       adsurl = {https://ui.adsabs.harvard.edu/abs/2018LPI....49.2570A}
}

@article{ReVelle1979453,
title = {A quasi-simple ablation model for large meteorite entry: {T}heory vs observations},
journal = {Journal of Atmospheric and Terrestrial Physics},
volume = {41},
number = {5},
pages = {453-473},
year = {1979},
issn = {0021-9169},
doi = {https://doi.org/10.1016/0021-9169(79)90071-0},
url = {https://www.sciencedirect.com/science/article/pii/0021916979900710},
author = {Douglas O ReVelle}
}

@article{Gritsevich2008fireballdynamics,
author = {Gritsevich, Maria},
year = {2008},
month = {02},
pages = {1-5},
title = {Identification of fireball dynamic parameters},
volume = {63},
journal = {Moscow University Mechanics Bulletin},
doi = {10.1007/s11971-008-1001-5}
}

@article{10.1093/mnrasl/slae065,
    author = {Peña-Asensio, E and Grèbol-Tomàs, P and Trigo-Rodríguez, J M and Ramírez-Moreta, P and Kresken, R},
    title = {The 18 {M}ay 2024 {I}berian superbolide from a sunskirting orbit: {USG} space sensors and ground-based independent observations},
    journal = {Monthly Notices of the Royal Astronomical Society: Letters},
    volume = {533},
    number = {1},
    pages = {L92-L99},
    year = {2024},
    month = {07},
    issn = {1745-3925},
    doi = {10.1093/mnrasl/slae065},
    url = {https://doi.org/10.1093/mnrasl/slae065},
    eprint = {https://academic.oup.com/mnrasl/article-pdf/533/1/L92/58736178/slae065.pdf},
}

@article{RN371,
   author = {Anderson, S. L. and Benedix, G. K. and Godel, B. and Alosius, R. M. L. and Krietsch, D. and Busemann, H. and Maden, C. and Friedrich, J. M. and McMonigal, L. R. and Welten, K. C. and Caffee, M. W. and Macke, R. J. and Cadogan, S. and Ryan, D. H. and Jourdan, F. and Mayers, C. and Laubenstein, M. and Greenwood, R. C. and Roberts, M. P. and Devillepoix, H. A. R. and Sansom, E. K. and Towner, M. C. and Cupák, M. and Bland, P. A. and Forman, L. V. and Fairweather, J. H. and Rogers, A. F. and Timms, N. E.},
   title = {The {A}rpu {K}uilpu meteorite: {I}n-depth characterization of an {H}5 chondrite delivered from a {J}upiter {F}amily {C}omet orbit},
   journal = {Meteoritics and Planetary Science},
   volume = {59},
   number = {11},
   pages = {3087-3110},
   ISSN = {1086-9379},
   doi = {10.1111/maps.14268},
   url = {<Go to ISI>://WOS:001325082400001},
   year = {2024},
   type = {Journal Article}
}

@article{RN3,
   author = {Anderson, S. L. and Towner, M. C. and Fairweather, J. and Bland, P. and Devillepoix, H. A. R. and Sansom, E. and Benedix, G. and Cupák, M.},
   title = {Successful Recovery of an Orbital Meteorite using drones and Machine Learning},
   journal = {Meteoritics and Planetary Science},
   volume = {57},
   ISSN = {1086-9379},
   url = {<Go to ISI>://WOS:000834630400017},
   year = {2022},
   type = {Journal Article}
}

@article{RN1705,
   author = {Andrade, M. and Docobo, J. A. and García-Guinea, J. and Campo, P. P. and Tapia, M. and Sánchez-Muñoz, L. and Villasante-Marcos, V. and Peña-Asensio, E. and Trigo-Rodríguez, J. M. and Ibáñez-Insa, J. and Campeny, M. and Llorca, J.},
   title = {The {T}raspena meteorite: {H}eliocentric orbit, atmospheric trajectory, strewn field, and petrography of a new {L}5 ordinary chondrite},
   journal = {Monthly Notices of the Royal Astronomical Society},
   volume = {518},
   number = {3},
   pages = {3850-3876},
   ISSN = {0035-8711},
   doi = {10.1093/mnras/stac2911},
   url = {<Go to ISI>://WOS:001051195600022},
   year = {2023},
   type = {Journal Article}
}

@article{RN1696,
   author = {Aoudjehane, H. C. and Agee, C. B. and Devillepoix, H. and Bouvier, A. and Benkhaldoun, Z. and Guennoun, M.},
   title = {Taghzout Meteorite: {T}he First Recovered Fall in {M}orocco Detected by {MOFID} cameras network.},
   journal = {Meteoritics and Planetary Science},
   volume = {59},
   pages = {A67-A67},
   ISSN = {1086-9379},
   url = {<Go to ISI>://WOS:001317679600068},
   year = {2024},
   type = {Journal Article}
}

@article{RN26,
   author = {Babadzhanov, P. B.},
   title = {Fragmentation and densities of meteoroids},
   journal = {Astronomy and Astrophysics},
   volume = {384},
   number = {1},
   pages = {317-321},
   ISSN = {0004-6361},
   doi = {10.1051/0004-6361:20020010},
   url = {<Go to ISI>://WOS:000174185500030},
   year = {2002},
   type = {Journal Article}
}

@article{Stulov1995aerodinamika,
  title={Aerodynamics of bolides},
  author={Stulov, VP and Mirskii, VN and Vislyi, AI},
  journal={Moscow: Science. Fizmatlit},
  year={1995}
}

@article{Boro2022a,
	author = {Borovička, J. and Spurný, P. and Shrbený, L. and Štork, R. and Kotková, L. and Fuchs, J. and Keclíková, J. and Zichová, H. and Mánek, J. and Váchová, P. and Macourková, I. and Svoreň, J. and Mucke, H.},
	title = {Data on 824 fireballs observed by the digital cameras of the {E}uropean {F}ireball {N}etwork in 2017–2018: {I}. {D}escription of the network, data reduction procedures, and the catalog},
	doi = {10.1051/0004-6361/202244184},
	url = {https://doi.org/10.1051/0004-6361/202244184},
	journal = {Astronomy and Astrophysics},
	year = 2022,
	volume = 667,
	pages = "A157",
}

@article{RN1354,
   author = {Barri, N. G. and Stulov, V. P.},
   title = {Peculiarities of the fragmentation of {B}enešov's bolide},
   journal = {Solar System Research},
   volume = {37},
   number = {4},
   pages = {302-305},
   ISSN = {0038-0946},
   doi = {10.1023/A:1025030331864},
   url = {<Go to ISI>://WOS:000185185700006},
   year = {2003},
   type = {Journal Article}
}

@article{RN1694,
   author = {Bischoff, A. and Barrat, J. A. and Bauer, K. and Burkhardt, C. and Busemann, H. and Ebert, S. and Gonsior, M. and Hakenmüller, J. and Haloda, J. and Harries, D. and Heinlein, D. and Hiesinger, H. and Hochleitner, R. and Hoffmann, V. and Kaliwoda, M. and Laubenstein, M. and Maden, C. and Meier, M. M. M. and Morlok, A. and Pack, A. and Ruf, A. and Schmitt-Kopplin, P. and Schönbächler, M. and Steele, R. C. J. and Spurný, P. and Wimmer, K.},
   title = {The {S}tubenberg meteorite: {A}n {LL}6 chondrite fragmental breccia recovered soon after precise prediction of the strewn field},
   journal = {Meteoritics and Planetary Science},
   volume = {52},
   number = {8},
   pages = {1683-1703},
   ISSN = {1086-9379},
   doi = {10.1111/maps.12883},
   url = {<Go to ISI>://WOS:000406869300010},
   year = {2017},
   type = {Journal Article}
}

@incollection{Jenniskens2024meteorshoweratlas,
title = {Chapter 4: {T}he Atlas: {O}verview and major showers},
editor = {Peter Jenniskens},
booktitle = {Atlas of Earth's Meteor Showers},
publisher = {Elsevier},
pages = {45-126},
year = {2024},
isbn = {978-0-443-23577-1},
doi = {https://doi.org/10.1016/B978-0-323-88447-1.00015-6},
url = {https://www.sciencedirect.com/science/article/pii/B9780323884471000156},
author = {Peter Jenniskens}
}

@article{Jenniskens2024solarsystempebbles,
title = {Properties of outer solar system pebbles during planetesimal formation from meteor observations},
journal = {Icarus},
volume = {423},
pages = {116229},
year = {2024},
issn = {0019-1035},
doi = {https://doi.org/10.1016/j.icarus.2024.116229},
url = {https://www.sciencedirect.com/science/article/pii/S0019103524002896},
author = {Peter Jenniskens and Paul R. Estrada and Stuart Pilorz and Peter S. Gural and Dave Samuels and Steve Rau and Timothy M.C. Abbott and Jim Albers and Scott Austin and Dan Avner and Jack W. Baggaley and Tim Beck and Solvay Blomquist and Mustafa Boyukata and Martin Breukers and Walt Cooney and Tim Cooper and Marcelo {De Cicco} and Hadrien Devillepoix and Eric Egland and Elize Fahl and Megan Gialluca and Bryant Grigsby and Toni Hanke and Barbara Harris and Steve Heathcote and Samantha Hemmelgarn and Andy Howell and Emmanuel Jehin and Carl Johannink and Luke Juneau and Erika Kisvarsanyi and Philip Mey and Nick Moskovitz and Mohammad Odeh and Brian Rachford and David Rollinson and James M. Scott and Martin C. Towner and Ozan Unsalan and Rynault {van Wyk} and Jeff Wood and James D. Wray and C. Pavao and Dante S. Lauretta}
}

@article{RN1691,
   author = {Bischoff, A. and Patzek, M. and Di Rocco, T. and Pack, A. and Stojic, A. and Berndt, J. and Peters, S.},
   title = {Saint-{P}ierre-le-{V}iger ({L}5-6) from asteroid 2023 {CX}1 recovered in the {N}ormandy, {F}rance 220 years after the historic fall of {L}'{A}igle ({L}6 breccia) in the neighborhood},
   journal = {Meteoritics and Planetary Science},
   volume = {58},
   number = {10},
   pages = {1385-1398},
   ISSN = {1086-9379},
   doi = {10.1111/maps.14074},
   url = {<Go to ISI>://WOS:001067531200001},
   year = {2023},
   type = {Journal Article}
}

@article{BlandArtemieva2006sizedistribution,
   author = {Bland, P. A. and Artemieva, N. A.},
   title = {The rate of small impacts on {E}arth},
   journal = {Meteoritics and Planetary Science},
   volume = {41},
   number = {4},
   pages = {607-631},
   ISSN = {1086-9379},
   doi = {10.1111/j.1945-5100.2006.tb00485.x},
   url = {<Go to ISI>://WOS:000237212300009},
   year = {2006},
   type = {Journal Article}
}

@article{RN552,
   author = {Boaca, I. and Gritsevich, M. and Birlan, M. and Nedelcu, A. and Boaca, T. and Colas, F. and Malgoyre, A. and Zanda, B. and Vernazza, P.},
   title = {Characterization of the Fireballs Detected by All-sky Cameras in {R}omania},
   journal = {Astrophysical Journal},
   volume = {936},
   number = {2},
   ISSN = {0004-637x},
   doi = {10.3847/1538-4357/ac8542},
   url = {<Go to ISI>://WOS:000852041000001},
   year = {2022},
   type = {Journal Article}
}

@ARTICLE{Boro2003moravkatrajectory,
       author = {Borovička, J. and {Spurn{\'y}}, P. and {Kalenda}, P. and {Tagliaferri}, E.},
        title = "{The {M}orávka meteorite fall: 1. {D}escription of the events and determination of the fireball trajectory and orbit from video records}",
      journal = {Meteoritics and Planetary Science},
         year = 2003,
        month = jul,
       volume = {38},
       number = {7},
        pages = {975-987},
          doi = {10.1111/j.1945-5100.2003.tb00293.x},
       adsurl = {https://ui.adsabs.harvard.edu/abs/2003M&PS...38..975B}
}

@article{BoroKalenda2003moravkafragmentation,
   author = {Borovička, J. and Kalenda, P.},
   title = {The {M}orávka meteorite fall: 4. {M}eteoroid dynamics and fragmentation in the atmosphere},
   journal = {Meteoritics and Planetary Science},
   volume = {38},
   number = {7},
   pages = {1023-1043},
   ISSN = {1086-9379},
   doi = {10.1111/j.1945-5100.2003.tb00296.x},
   url = {<Go to ISI>://WOS:000186306000006},
   year = {2003},
   type = {Journal Article}
}

@article{RN113,
   author = {Borovička, J. and Koten, P. and Spurný, P. and Bocek, J. and Štork, R.},
   title = {A survey of meteor spectra and orbits: {E}vidence for three populations of {N}a-free meteoroids},
   journal = {Icarus},
   volume = {174},
   number = {1},
   pages = {15-30},
   ISSN = {0019-1035},
   doi = {10.1016/j.icarus.2004.09.011},
   url = {<Go to ISI>://WOS:000228136900002},
   year = {2005},
   type = {Journal Article}
}

@article{RN127,
   author = {Borovička, J. and Popova, O. and Spurný, P.},
   title = {The {M}aribo {CM}2 meteorite fall: {S}urvival of weak material at high entry speed},
   journal = {Meteoritics and Planetary Science},
   volume = {54},
   number = {5},
   pages = {1024-1041},
   ISSN = {1086-9379},
   doi = {10.1111/maps.13259},
   url = {<Go to ISI>://WOS:000468026900003},
   year = {2019},
   type = {Journal Article}
}

@article{Boro2015krizevci,
   author = {Borovička, J. and Spurný, P. and Šegon, D. and Andreic, Z. and Kac, J. and Korlevic, K. and Atanackov, J. and Kladnik, G. and Mucke, H. and Vida, D. and Novoselnik, F.},
   title = {The instrumentally recorded fall of the {K}riževci meteorite, {C}roatia, {F}ebruary 4, 2011},
   journal = {Meteoritics and Planetary Science},
   volume = {50},
   number = {7},
   pages = {1244-1259},
   ISSN = {1086-9379},
   doi = {10.1111/maps.12469},
   url = {<Go to ISI>://WOS:000358126000006},
   year = {2015},
   type = {Journal Article}
}

@article{RN95,
   author = {Borovička, J. and Spurný, P. and Shrbený, L.},
   title = {Two Strengths of Ordinary Chondritic Meteoroids as Derived from Their Atmospheric Fragmentation Modeling},
   journal = {Astronomical Journal},
   volume = {160},
   number = {1},
   ISSN = {0004-6256},
   doi = {10.3847/1538-3881/ab9608},
   url = {<Go to ISI>://WOS:000545671000001},
   year = {2020},
   type = {Journal Article}
}

@article{RN158,
   author = {Brown, P. and McCausland, P. J. A. and Fries, M. and Silber, E. and Edwards, W. N. and Wong, D. K. and Weryk, R. J. and Fries, J. and Krzeminski, Z.},
   title = {The fall of the {G}rimsby meteorite: {I}. {F}ireball dynamics and orbit from radar, video, and infrasound records},
   journal = {Meteoritics and Planetary Science},
   volume = {46},
   number = {3},
   pages = {339-363},
   ISSN = {1086-9379},
   doi = {10.1111/j.1945-5100.2010.01167.x},
   url = {<Go to ISI>://WOS:000288500900001},
   year = {2011},
   type = {Journal Article}
}

@article{RN1745,
   author = {Brown, P. and Pack, D. and Edwards, W. N. and ReVelle, D. O. and Yoo, B. B. and Spalding, R. E. and Tagliaferri, E.},
   title = {The orbit, atmospheric dynamics, and initial mass of the {P}ark {F}orest meteorite},
   journal = {Meteoritics and Planetary Science},
   volume = {39},
   number = {11},
   pages = {1781-1796},
   ISSN = {1086-9379},
   doi = {10.1111/j.1945-5100.2004.tb00075.x},
   url = {<Go to ISI>://WOS:000226211300002},
   year = {2004},
   type = {Journal Article}
}

@article{RN1702,
   author = {Brown, P. G. and Hildebrand, A. R. and Zolensky, M. E. and Grady, M. and Clayton, R. N. and Mayeda, T. K. and Tagliaferri, E. and Spalding, R. and MacRae, N. D. and Hoffman, E. L. and Mittlefehldt, D. W. and Wacker, J. F. and Bird, J. A. and Campbell, M. D. and Carpenter, R. and Gingerich, H. and Glatiotis, M. and Greiner, E. and Mazur, M. J. and McCausland, P. J. A. and Plotkin, H. and Mazur, T. R.},
   title = {The fall, recovery, orbit, and composition of the {T}agish {L}ake meteorite: {A} new type of carbonaceous chondrite},
   journal = {Science},
   volume = {290},
   number = {5490},
   pages = {320-325},
   ISSN = {0036-8075},
   doi = {10.1126/science.290.5490.320},
   url = {<Go to ISI>://WOS:000089818100039},
   year = {2000},
   type = {Journal Article}
}

@article{RN162,
   author = {Brown, P. G. and McCausland, P. J. A. and Hildebrand, A. R. and Hanton, L. T. J. and Eckart, L. M. and Busemann, H. and Krietsch, D. and Maden, C. and Welten, K. and Caffee, M. W. and Laubenstein, M. and Vida, D. and Ciceri, F. and Silber, E. and Herd, C. D. K. and Hill, P. and Devillepoix, H. and Sansom, E. K. and Cupák, M. and Anderson, S. and Flemming, R. L. and Nelson, A. J. and Mazur, M. and Moser, D. E. and Cooke, W. J. and Hladiuk, D. and Malecic, B. and Prtenjak, M. T. and Nowell, R. and Consortium, Golden Meteorite},
   title = {The {G}olden meteorite fall: {F}ireball trajectory, orbit, and meteorite characterization},
   journal = {Meteoritics and Planetary Science},
   volume = {58},
   number = {12},
   pages = {1773-1807},
   ISSN = {1086-9379},
   doi = {10.1111/maps.14100},
   url = {<Go to ISI>://WOS:001109370200001},
   year = {2023},
   type = {Journal Article}
}

@article{RN1700,
   author = {Brown, P. G. and ReVelle, D. O. and Tagliaferri, E. and Hildebrand, A. R.},
   title = {An entry model for the {T}agish {L}ake fireball using seismic, satellite and infrasound records},
   journal = {Meteoritics and Planetary Science},
   volume = {37},
   number = {5},
   pages = {661-675},
   ISSN = {1086-9379},
   doi = {10.1111/j.1945-5100.2002.tb00846.x},
   url = {<Go to ISI>://WOS:000175971600004},
   year = {2002},
   type = {Journal Article}
}

@article{RN164,
   author = {Brown, P. G. and Vida, D. and Moser, D. E. and Granvik, M. and Koshak, W. J. and Chu, D. and Steckloff, J. and Licata, A. and Hariri, S. and Mason, J. and Mazur, M. and Cooke, W. and Krzeminski, Z.},
   title = {The {H}amburg meteorite fall: {F}ireball trajectory, orbit, and dynamics},
   journal = {Meteoritics and Planetary Science},
   volume = {54},
   number = {9},
   pages = {2027-2045},
   ISSN = {1086-9379},
   doi = {10.1111/maps.13368},
   url = {<Go to ISI>://WOS:000481038800001},
   year = {2019},
   type = {Journal Article}
}

@ARTICLE{PecinaCeplecha1983,
       author = {{Pecina}, P. and {Ceplecha}, Z.},
        title = "{New Aspects in Single-Body meteor Physics}",
      journal = {Bulletin of the Astronomical Institutes of Czechoslovakia},
         year = 1983,
        month = mar,
       volume = {34},
        pages = {102},
       adsurl = {https://ui.adsabs.harvard.edu/abs/1983BAICz..34..102P}
}

@article{Boro2007,
	author = {Borovička, J. and Spurný, P. and Koten, P.},
	title = {Atmospheric deceleration and light curves of {D}raconid meteors and implications for the structure of cometary dust},
	doi= {10.1051/0004-6361:20078131},
	url= {https://doi.org/10.1051/0004-6361:20078131},
	journal = {Astronomy and Astrophysics},
	year = 2007,
	volume = 473,
	number = 2,
	pages = "661-672",
}

@article{Buccongello2024,
   author = {Buccongello, N. and Brown, P. G. and Vida, D. and Pinhas, A.},
   title = {A physical survey of meteoroid streams: {C}omparing cometary reservoirs},
   journal = {Icarus},
   volume = {410},
   ISSN = {0019-1035},
   doi = {10.1016/j.icarus.2023.115907},
   url = {<Go to ISI>://WOS:001135265800001},
   year = {2024},
   type = {Journal Article}
}

@article{RN365,
   author = {Ceplecha, Z.},
   title = {Ablation and Shape-Density Coefficients in Meteors},
   journal = {Bulletin of the Astronomical Institutes of Czechoslovakia},
   volume = {26},
   number = {4},
   pages = {242-248},
   ISSN = {0004-6248},
   url = {<Go to ISI>://WOS:A1975AM19700010},
   year = {1975},
   type = {Journal Article}
}

@article{RN357,
   author = {Ceplecha, Z. and Borovička, J. and Elford, W. G. and ReVelle, D. O. and Hawkes, R. L. and Porubčan, V. and Simek, M.},
   title = {Meteor phenomena and bodies},
   journal = {Space Science Reviews},
   volume = {84},
   number = {3-4},
   pages = {327-471},
   ISSN = {0038-6308},
   doi = {10.1023/A:1005069928850},
   url = {<Go to ISI>://WOS:000076704800001},
   year = {1998},
   type = {Journal Article}
}

@article{RN1229,
   author = {Devillepoix, H. A. R. and Anderson, S. and Sansom, E. K. and Lagain, A. and Towner, M. C. and Bland, P. A. and Howie, R. M. and Cupák, M. and Benedix, G. K. and Forman, L. V. and Shober, P. and Hartig, B. A. D.},
   title = {Madura {C}ave: {A} New Meteorite Fall Delivered from an {A}ten Orbit},
   journal = {Meteoritics and Planetary Science},
   volume = {56},
   ISSN = {1086-9379},
   url = {<Go to ISI>://WOS:000684014300063},
   year = {2021},
   type = {Journal Article}
}

@article{RN436,
   author = {Devillepoix, H. A. R. and Bland, P. A. and Towner, M. C. and Sansom, E. K. and Howie, R. M. and Cupák, M. and Benedix, G. K. and Jansen-Sturgeon, T. and Hartig, B. A. D. and Cox, M. A. and Paxman, J. P.},
   title = {Fall and Recovery of the {D}ingle {D}ell Meteorite},
   journal = {Meteoritics and Planetary Science},
   volume = {52},
   pages = {A69-A69},
   ISSN = {1086-9379},
   url = {<Go to ISI>://WOS:000418552900070},
   year = {2017},
   type = {Journal Article}
}

@article{RN166,
   author = {Devillepoix, H. A. R. and Cupák, M. and Bland, P. A. and Sansom, E. K. and Towner, M. C. and Howie, R. M. and Hartig, B. A. D. and Jansen-Sturgeon, T. and Shober, P. M. and Anderson, S. L. and Benedix, G. K. and Busan, D. and Sayers, R. and Jenniskens, P. and Albers, J. and Herd, C. D. K. and Hill, P. J. A. and Brown, P. G. and Krzeminski, Z. and Osinski, G. R. and Aoudjehane, H. C. and Benkhaldoun, Z. and Jabiri, A. and Guennoun, M. and Barka, A. and Darhmaoui, H. and Daly, L. and Collins, G. S. and McMullan, S. and Suttle, M. D. and Ireland, T. and Bonning, G. and Baeza, L. and Alrefay, T. Y. and Horner, J. and Swindle, T. D. and Hergenrother, C. W. and Fries, M. D. and Tomkins, A. and Langendam, A. and Rushmer, T. and O'Neill, C. and Janches, D. and Hormaechea, J. L. and Shaw, C. and Young, J. S. and Alexander, M. and Mardon, A. D. and Tate, J. R.},
   title = {A {G}lobal {F}ireball {O}bservatory},
   journal = {Planetary and Space Science},
   volume = {191},
   ISSN = {0032-0633},
   doi = {10.1016/j.pss.2020.105036},
   url = {<Go to ISI>://WOS:000561217400004},
   year = {2020},
   type = {Journal Article}
}

@article{RN432,
   author = {Devillepoix, H. A. R. and Sansom, E. K. and Bland, P. A. and Towner, M. C. and Cupák, M. and Howie, R. M. and Jansen-Sturgeon, T. and Cox, M. A. and Hartig, B. A. D. and Benedix, G. K. and Paxman, J. P.},
   title = {The {D}ingle {D}ell meteorite: {A} {H}alloween treat from the {M}ain {B}elt},
   journal = {Meteoritics and Planetary Science},
   volume = {53},
   number = {10},
   pages = {2212-2227},
   ISSN = {1086-9379},
   doi = {10.1111/maps.13142},
   url = {<Go to ISI>://WOS:000446173000012},
   year = {2018},
   type = {Journal Article}
}

@article{RN5,
   author = {Devillepoix, H. A. R. and Sansom, E. K. and Shober, P. and Anderson, S. L. and Towner, M. C. and Lagain, A. and Cupák, M. and Bland, P. A. and Howie, R. M. and Jansen-Sturgeon, T. and Hartig, B. A. D. and Sokolowski, M. and Benedix, G. and Forman, L.},
   title = {Trajectory, recovery, and orbital history of the {M}adura {C}ave meteorite},
   journal = {Meteoritics and Planetary Science},
   volume = {57},
   number = {7},
   pages = {1328-1338},
   ISSN = {1086-9379},
   doi = {10.1111/maps.13820},
   url = {<Go to ISI>://WOS:000804475300001},
   year = {2022},
   type = {Journal Article}
}

@article{RN1615,
   author = {Dumitru, B. A. and Birlan, M. and Popescu, M. and Nedelcu, D. A.},
   title = {Association between meteor showers and asteroids using multivariate criteria},
   journal = {Astronomy and Astrophysics},
   volume = {607},
   ISSN = {0004-6361},
   doi = {10.1051/0004-6361/201730813},
   url = {<Go to ISI>://WOS:000414180200005},
   year = {2017},
   type = {Journal Article}
}

@article{Gardiol2021cavezzo,
   author = {Gardiol, D. and Barghini, D. and Buzzoni, A. and Carbognani, A. and Di Carlo, M. and Di Martino, M. and Knapic, C. and Londero, E. and Pratesi, G. and Rasetti, S. and Riva, W. and Salerno, R. and Stirpe, G. M. and Valsecchi, G. B. and Volpicelli, C. A. and Zorba, S. and Colas, F. and Zanda, B. and Bouley, S. and Jeanne, S. and Malgoyre, A. and Birlan, M. and Blanpain, C. and Gattacceca, J. and Lecubin, J. and Marmo, C. and Rault, J. L. and Vaubaillon, J. and Vernazza, P. and Affaticati, F. and Albani, M. and Andreis, A. and Ascione, G. and Avoscan, T. and Bacci, P. and Baldini, R. and Balestrero, A. and Basso, S. and Bellitto, R. and Belluso, M. and Benna, C. and Bernardi, F. and Bertaina, M. E. and Betti, L. and Bonino, R. and Boros, K. and Bussi, A. and Carli, C. and Carriero, T. and Cascone, E. and Cattaneo, C. and Cellino, A. and Colombetti, P. and Colombi, E. and Costa, M. and Cremonese, G. and Cricchio, D. and D'Agostino, G. and D'Elia, M. and De Maio, M. and Demaria, P. and Di Dato, A. and Di Luca, R. and Federici, F. and Gagliarducci, V. and Gerardi, A. and Giuli, G. and Guidetti, D. and Interrante, G. and Lazzarin, M. and Lera, S. and Leto, G. and Licchelli, D. and Lippolis, F. and Manca, F. and Mancuso, S. and Mannucci, F. and Masi, R. and Masiero, S. and Meucci, S. and Misiano, A. and Cecchi, V. M. and Molinari, E. and Monari, J. and Montemaggi, M. and Montesarchio, M. and Monti, G. and Morini, P. and Nastasi, A. and Pace, E. and Pardini, R. and Pavone, M. and Pegoraro, A. and Pietronave, S. and Pisanu, T. and Pugno, N. and Repetti, U. and Rigoni, M. and Rizzi, N. and Romeni, C. and others },
   title = {Cavezzo, the first {I}talian meteorite recovered by the {PRISMA} fireball network: {O}rbit, trajectory, and strewn-field},
   journal = {Monthly Notices of the Royal Astronomical Society},
   volume = {501},
   number = {1},
   pages = {1215-1227},
   ISSN = {0035-8711},
   doi = {10.1093/mnras/staa3646},
   url = {<Go to ISI>://WOS:000608474800093},
   year = {2021},
   type = {Journal Article}
}

@article{RN455,
   author = {Girin, O. G.},
   title = {A hydrodynamic mechanism of meteor ablation: {T}he melt-spraying model},
   journal = {Astronomy and Astrophysics},
   volume = {606},
   ISSN = {0004-6361},
   doi = {10.1051/0004-6361/201629560},
   url = {<Go to ISI>://WOS:000412876800002},
   year = {2017},
   type = {Journal Article}
}

@article{RN467,
   author = {Greenberg, J. M.},
   title = {Making a comet nucleus},
   journal = {Astronomy and Astrophysics},
   volume = {330},
   number = {1},
   pages = {375-380},
   ISSN = {1432-0746},
   url = {<Go to ISI>://WOS:000071970100050},
   year = {1998},
   type = {Journal Article}
}

@article{RN518,
   author = {Greenwood, R. C. and Burbine, T. H. and Franchi, I. A.},
   title = {Linking asteroids and meteorites to the primordial planetesimal population},
   journal = {Geochimica Et Cosmochimica Acta},
   volume = {277},
   pages = {377-406},
   ISSN = {0016-7037},
   url = {<Go to ISI>://WOS:000530720600020},
   year = {2020},
   type = {Journal Article}
}

@article{LeeBland2004meteoriteweathering,
title = {Mechanisms of weathering of meteorites recovered from hot and cold deserts and the formation of phyllosilicates},
journal = {Geochimica et Cosmochimica Acta},
volume = {68},
number = {4},
pages = {893-916},
year = {2004},
issn = {0016-7037},
doi = {https://doi.org/10.1016/S0016-7037(03)00486-1},
url = {https://www.sciencedirect.com/science/article/pii/S0016703703004861},
author = {Martin R. Lee and Philip A. Bland}
}

@article{RN538,
   author = {Gritsevich, M. and Moilanen, J. and Visuri, J. and Meier, M. M. M. and Maden, C. and Oberst, J. and Heinlein, D. and Flohrer, J. and Castro-Tirado, A. J. and Delgado-García, J. and Koeberl, C. and Ferrière, L. and Brandstätter, F. and Povinec, P. P. and Sykora, I. and Schweidler, F.},
   title = {The fireball of {N}ovember 24, 1970, as the most probable source of the {I}schgl meteorite},
   journal = {Meteoritics and Planetary Science},
   volume = {59},
   number = {7},
   pages = {1658-1691},
   ISSN = {1086-9379},
   doi = {10.1111/maps.14173},
   url = {<Go to ISI>://WOS:001230608200001},
   year = {2024},
   type = {Journal Article}
}

@misc{Moilanen2026adalen,
      title={The First Instrumentally Documented Fall of an Iron Meteorite: {A}tmospheric trajectory and ground impact}, 
      author={Jarmo Moilanen and Maria Gritsevich and Jaakko Visuri},
      year={2026},
      eprint={2602.15440},
      archivePrefix={arXiv},
      primaryClass={astro-ph.EP},
      url={https://arxiv.org/abs/2602.15440}, 
}

@article{Gritsevich2008,
   author = {Gritsevich, M.},
   title = {The {P}říbram, {L}ost {C}ity, {I}nnisfree, and {N}euschwanstein falls: {A}n analysis of the atmospheric trajectories},
   journal = {Solar System Research},
   volume = {42},
   number = {5},
   pages = {372-390},
   ISSN = {0038-0946},
   doi = {10.1134/S003809460805002x},
   url = {<Go to ISI>://WOS:000259394700002},
   year = {2008},
   type = {Journal Article}
}

@article{RN691,
   author = {Gritsevich, M.},
   title = {Determination of parameters of meteor bodies based on flight observational data},
   journal = {Advances in Space Research},
   volume = {44},
   number = {3},
   pages = {323-334},
   ISSN = {0273-1177},
   doi = {10.1016/j.asr.2009.03.030},
   url = {<Go to ISI>://WOS:000268360900004},
   year = {2009},
   type = {Journal Article}
}

@article{RN696,
   author = {Gritsevich, M. and Popelenskaya, N. V.},
   title = {Meteor and fireball trajectories for high values of the mass loss parameter},
   journal = {Doklady Physics},
   volume = {53},
   number = {2},
   pages = {88-92},
   ISSN = {1028-3358},
   doi = {10.1134/S1028335808020092},
   url = {<Go to ISI>://WOS:000253723900009},
   year = {2008},
   type = {Journal Article}
}

@article{RN694,
   author = {Gritsevich, M. and Stulov, V. P.},
   title = {A model of the motion of the {N}euschwanstein bolide in the atmosphere},
   journal = {Solar System Research},
   volume = {42},
   number = {2},
   pages = {118-123},
   ISSN = {0038-0946},
   doi = {10.1134/S0038094608020032},
   url = {<Go to ISI>://WOS:000254840000003},
   year = {2008},
   type = {Journal Article}
}

@ARTICLE{Gritsevich2012impactconsequences,
       author = {{Gritsevich}, M.~I. and {Stulov}, V.~P. and {Turchak}, L.~I.},
        title = "{Consequences of collisions of natural cosmic bodies with the {E}arth's atmosphere and surface}",
      journal = {Cosmic Research},
         year = 2012,
        month = feb,
       volume = {50},
       number = {1},
        pages = {56-64},
          doi = {10.1134/S0010952512010017},
       adsurl = {https://ui.adsabs.harvard.edu/abs/2012CosRe..50...56G}
}

@article{RN1697,
   author = {Guennoun, M. and Devillepoix, H. A. R. and Cupák, M. and Benkhaldoun, Z. and Aoudjehane, H. C. and Bouvier, A.},
   title = {Fall Area Calculation and Association of Meteor Orbits with Parent Bodies: {A} Case Study Using the {MOFID} Network for {T}aghzout Meteorite Fall, {M}orocco.},
   journal = {Meteoritics and Planetary Science},
   volume = {59},
   pages = {A170-A170},
   ISSN = {1086-9379},
   url = {<Go to ISI>://WOS:001317679600171},
   year = {2024},
   type = {Journal Article}
}

@article{RN1607,
   author = {Halliday, I.},
   title = {Geminid Fireballs and the Peculiar Asteroid 3200 {P}haethon},
   journal = {Icarus},
   volume = {76},
   number = {2},
   pages = {279-294},
   ISSN = {0019-1035},
   doi = {10.1016/0019-1035(88)90073-5},
   url = {<Go to ISI>://WOS:A1988Q976500005},
   year = {1988},
   type = {Journal Article}
}

@article{Gritsevich2007alphabeta,
author = {Gritsevich, M.},
year = {2007},
month = {01},
pages = {509-514},
title = {Approximation of the observed motion of bolides by the analytical solution of the equations of meteor physics},
volume = {41},
journal = {Solar System Research},
doi = {10.1134/S003809460706007X}
}

@article{Silber2018meteorshockwaves,
title = {Physics of meteor generated shock waves in the {E}arth’s atmosphere: {A} review},
journal = {Advances in Space Research},
volume = {62},
number = {3},
pages = {489-532},
year = {2018},
issn = {0273-1177},
doi = {https://doi.org/10.1016/j.asr.2018.05.010},
url = {https://www.sciencedirect.com/science/article/pii/S027311771830406X},
author = {Elizabeth A. Silber and Mark Boslough and Wayne K. Hocking and Maria Gritsevich and Rodney W. Whitaker}
}

@article{RN1753,
   author = {Jenniskens, P. and Devillepoix, H. A. R.},
   title = {Review of asteroid, meteor, and meteorite-type links},
   journal = {Meteoritics and Planetary Science},
   volume = {60},
   number = {4},
   pages = {928-973},
   ISSN = {1086-9379},
   doi = {10.1111/maps.14321},
   url = {<Go to ISI>://WOS:001464450200001},
   year = {2025},
   type = {Journal Article}
}

@article{Ceplecha2007tagishlakefragmentation,
author = {Ceplecha, Zdeněk},
title = {Fragmentation model analysis of the observed atmospheric trajectory of the {T}agish {L}ake fireball},
journal = {Meteoritics and Planetary Science},
volume = {42},
number = {2},
pages = {185-189},
doi = {https://doi.org/10.1111/j.1945-5100.2007.tb00226.x},
url = {https://onlinelibrary.wiley.com/doi/abs/10.1111/j.1945-5100.2007.tb00226.x},
eprint = {https://onlinelibrary.wiley.com/doi/pdf/10.1111/j.1945-5100.2007.tb00226.x},
year = {2007}
}

@article{CeplechaReVelle2005fragmentationmodel,
author = {Ceplecha, Zdeněk and ReVelle, Douglas O.},
title = {Fragmentation model of meteoroid motion, mass loss, and radiation in the atmosphere},
journal = {Meteoritics and Planetary Science},
volume = {40},
number = {1},
pages = {35-54},
doi = {https://doi.org/10.1111/j.1945-5100.2005.tb00363.x},
url = {https://onlinelibrary.wiley.com/doi/abs/10.1111/j.1945-5100.2005.tb00363.x},
eprint = {https://onlinelibrary.wiley.com/doi/pdf/10.1111/j.1945-5100.2005.tb00363.x},
year = {2005}
}

@INPROCEEDINGS{Ceplecha1994meteoroidproperties,
       author = {{Ceplecha}, Z.},
        title = "{Meteoroid Properties from Photographic Records of Meteors and Fireballs}",
    booktitle = {Asteroids, Comets, Meteors 1993},
         year = 1994,
       editor = {{Milani}, Andrea and {di Martino}, Michel and {Cellino}, A.},
       series = {IAU Symposium},
       volume = {160},
        month = jan,
        pages = {343},
       adsurl = {https://ui.adsabs.harvard.edu/abs/1994IAUS..160..343C}
}

@article{Zappatini2026alkhadhaf,
author = {Zappatini, Anna and Gnos, Edwin and Hofmann, Beda A. and Eggenberger, Urs and Kruttasch, Pascal M. and Gfeller, Frank and Tauseef, Mohammad and Leya, Ingo and Devillepoix, Hadrien A. R. and Sansom, Eleanor K. and Cupák, Martin and Deam, Sophie E. and Stevenson, Thomas W. C. and Jenniskens, Peter and Lindemann, Sebastian and Booz, Beat and Al-Muati, Muati. S. and Al-Zakwani, Abdulmunaim A. and Al-Ghafri, Hussain A.},
title = {Al-{K}hadhaf: {T}he first camera-observed ({H}5–6) meteorite fall from {O}man},
journal = {Meteoritics \& Planetary Science},
year = {2026},
doi = {https://doi.org/10.1111/maps.70110},
url = {https://onlinelibrary.wiley.com/doi/abs/10.1111/maps.70110},
eprint = {https://onlinelibrary.wiley.com/doi/pdf/10.1111/maps.70110}
}

@article{Bouquet2014orbitaldetectionsimulation,
title = {Simulation of the capabilities of an orbiter for monitoring the entry of interplanetary matter into the terrestrial atmosphere},
journal = {Planetary and Space Science},
volume = {103},
pages = {238-249},
year = {2014},
issn = {0032-0633},
doi = {https://doi.org/10.1016/j.pss.2014.09.001},
url = {https://www.sciencedirect.com/science/article/pii/S0032063314002785},
author = {Alexis Bouquet and David Baratoux and Jérémie Vaubaillon and Maria I. Gritsevich and David Mimoun and Olivier Mousis and Sylvain Bouley}
}

@article{GritsevichKoschny2011luminousefficiency,
title = {Constraining the luminous efficiency of meteors},
journal = {Icarus},
volume = {212},
number = {2},
pages = {877-884},
year = {2011},
issn = {0019-1035},
doi = {https://doi.org/10.1016/j.icarus.2011.01.033},
url = {https://www.sciencedirect.com/science/article/pii/S0019103511000443},
author = {Maria Gritsevich and Detlef Koschny}
}

@article{RN1612,
   author = {Jenniskens, P. and Vaubaillon, J.},
   title = {Minor Planet 2002 {EX}12
 (=169{P}/{NEAT}) and the {A}lpha {C}apricornid Shower},
   journal = {Astronomical Journal},
   volume = {139},
   number = {5},
   pages = {1822-1830},
   ISSN = {0004-6256},
   doi = {10.1088/0004-6256/139/5/1822},
   url = {<Go to ISI>://WOS:000276513700007},
   year = {2010},
   type = {Journal Article}
}

@INPROCEEDINGS{Kuznetsova2014kosicetrajectory,
       author = {{Kuznetsova}, Daria and {Gritsevich}, Maria and {Vinnikov}, Vladimir},
        title = "{The {K}ošice meteoroid investigation: {F}rom trajectory data to analytic model}",
    booktitle = {Proceedings of the International Meteor Conference, Giron, France, 18-21 September 2014},
         year = 2014,
       editor = {{Rault}, J. -L. and {Roggemans}, P.},
        month = feb,
        pages = {178-181},
       adsurl = {https://ui.adsabs.harvard.edu/abs/2014pim4.conf..178K}
}

@article{RN865,
   author = {Jewitt, D. and Li, J.},
   title = {Activity in {G}eminid Parent (3200) {P}haethon},
   journal = {Astronomical Journal},
   volume = {140},
   number = {5},
   pages = {1519-1527},
   ISSN = {0004-6256},
   doi = {10.1088/0004-6256/140/5/1519},
   url = {<Go to ISI>://WOS:000283055400035},
   year = {2010},
   type = {Journal Article}
}

@article{RN862,
   author = {Jewitt, D. and Mutchler, M. and Agarwal, J. and Li, J.},
   title = {Observations of 3200 {P}haethon at Closest Approach},
   journal = {Astronomical Journal},
   volume = {156},
   number = {5},
   ISSN = {0004-6256},
   doi = {10.3847/1538-3881/aae51f},
   url = {<Go to ISI>://WOS:000449182800006},
   year = {2018},
   type = {Journal Article}
}

@article{RN1716,
   author = {King, A. J. and Daly, L. and Rowe, J. and Joy, K. H. and Greenwood, R. C. and Devillepoix, H. A. R. and Suttle, M. D. and Chan, Q. H. S. and Russell, S. S. and Bates, H. C. and Bryson, J. F. J. and Clay, P. L. and Vida, D. and Lee, M. R. and O'Brien, A. and Hallis, L. J. and Stephen, N. R. and Tartèse, R. and Sansom, E. K. and Towner, M. C. and Cupák, M. and Shober, P. M. and Bland, P. A. and Findlay, R. and Franchi, I. A. and Verchovsky, A. B. and Abernethy, F. A. J. and Grady, M. M. and Floyd, C. J. and Van Ginneken, M. and Bridges, J. and Hicks, L. J. and Jones, R. H. and Mitchell, J. T. and Genge, M. J. and Jenkins, L. and Martin, P. E. and Sephton, M. A. and Watson, J. S. and Salge, T. and Shirley, K. A. and Curtis, R. J. and Warren, T. J. and Bowles, N. E. and Stuart, F. M. and Di Nicola, L. and Györe, D. and Boyce, A. J. and Shaw, K. M. M. and Elliott, T. and Steele, R. C. J. and Povinec, P. and Laubenstein, M. and Sanderson, D. and Cresswell, A. and Jull, A. J. T. and Sykora, I. and Sridhar, S. and Harrison, R. J. and Willcocks, F. M. and Harrison, C. S. and Hallatt, D. and Wozniakiewicz, P. J. and Burchell, M. J. and Alesbrook, L. S. and Dignam, A. and Almeida, N. and Smith, C. L. and Clark, B. and Humphreys-Williams, E. R. and Schofield, P. F. and Cornwell, L. T. and Spathis, V. and Morgan, G. H. and Perkins, M. J. and Kacerek, R. and Campbell-Burns, P. and Colas, F. and Zanda, B. and Vernazza, P. and Bouley, S. and Jeanne, S. and Hankey, M. and Collins, G. S. and Young, J. S. and Shaw, C. and Horak, J. and Jones, D. and James, N. and Bosley, S. and Shuttleworth, A. and Dickinson, P. and McMullan, I. and Robson, D. and Smedley, A. R. D. and Stanley, B. and Bassom, R. and McIntyre, M. and Suttle, A. A. and Fleet, R. and others },
   title = {The {W}inchcombe meteorite, a unique and pristine witness from the outer {S}olar {S}ystem},
   journal = {Science Advances},
   volume = {8},
   number = {46},
   ISSN = {2375-2548},
   doi = {10.1126/sciadv.abq3925},
   url = {<Go to ISI>://WOS:000945384000002},
   year = {2022},
   type = {Journal Article}
}

@article{RN543,
   author = {Kyrylenko, I. and Golubov, O. and Slyusarev, I. and Visuri, J. and Gritsevich, M. and Krugly, Y. N. and Belskaya, I. and Shevchenko, V. G.},
   title = {The First Instrumentally Documented Fall of an Iron Meteorite: {O}rbit and Possible Origin},
   journal = {Astrophysical Journal},
   volume = {953},
   number = {1},
   ISSN = {0004-637x},
   doi = {10.3847/1538-4357/acdc21},
   url = {<Go to ISI>://WOS:001038721600001},
   year = {2023},
   type = {Journal Article}
}

@article{RN658,
   author = {Lyytinen, E. and Gritsevich, M.},
   title = {Implications of the atmospheric density profile in the processing of fireball observations},
   journal = {Planetary and Space Science},
   volume = {120},
   pages = {35-42},
   ISSN = {0032-0633},
   doi = {10.1016/j.pss.2015.10.012},
   url = {<Go to ISI>://WOS:000370101000004},
   year = {2016},
   type = {Journal Article}
}

@article{RN873,
   author = {Matlovič, P. and Pisarciková, A. and Pazderová, V. and Loehle, S. and Tóth, J. and Ferrière, L. and Cermák, P. and Leiser, D. and Vaubaillon, J. and Ravichandran, R.},
   title = {Spectral properties of ablating meteorite samples for improved meteoroid composition diagnostics},
   journal = {Astronomy and Astrophysics},
   volume = {689},
   ISSN = {0004-6361},
   doi = {10.1051/0004-6361/202450913},
   url = {<Go to ISI>://WOS:001318793400011},
   year = {2024},
   type = {Journal Article}
}

@article{RN890,
   author = {Matlovič, P. and Tóth, J. and Rudawska, R. and Kornos, L. and Pisarcíková, A.},
   title = {Spectral and orbital survey of medium-sized meteoroids},
   journal = {Astronomy and Astrophysics},
   volume = {629},
   ISSN = {0004-6361},
   doi = {10.1051/0004-6361/201936093},
   url = {<Go to ISI>://WOS:000484658100004},
   year = {2019},
   type = {Journal Article}
}

@article{McFadden2024fireballacoustics,
title = {A comparison of fireball luminous efficiency models using acoustic records},
journal = {Icarus},
volume = {422},
pages = {116250},
year = {2024},
issn = {0019-1035},
doi = {https://doi.org/10.1016/j.icarus.2024.116250},
url = {https://www.sciencedirect.com/science/article/pii/S0019103524003105},
author = {{McFadden, L.}, {Brown, P. G.} & {Vida, D.}}
}

@article{RN1711,
   author = {McMullan, S. and Vida, D. and Devillepoix, H. A. R. and Rowe, J. and Daly, L. and King, A. J. and Cupák, M. and Howie, R. M. and Sansom, E. K. and Shober, P. and Towner, M. C. and Anderson, S. and McFadden, L. and Horák, J. and Smedley, A. R. D. and Joy, K. H. and Shuttleworth, A. and Colas, F. and Zanda, B. and O'Brien, A. C. and McMullan, I. and Shaw, C. and Suttle, A. and Suttle, M. D. and Young, J. S. and Campbell-Burns, P. and Kacerek, R. and Bassom, R. and Bosley, S. and Fleet, R. and Jones, D. and McIntyre, M. and James, N. and Robson, D. and Dickinson, P. and Bland, P. A. and Collins, G. S.},
   title = {The {W}inchcombe fireball: {T}hat lucky survivor},
   journal = {Meteoritics and Planetary Science},
   volume = {59},
   number = {5},
   pages = {927-947},
   ISSN = {1086-9379},
   doi = {10.1111/maps.13977},
   url = {<Go to ISI>://WOS:001218785100010},
   year = {2024},
   type = {Journal Article}
}

@Inbook{CampbellBrown2005morp,
author="Campbell-Brown, M. D. and Hildebrand, A.",
editor="Hawkes, Robert and Mann, Ingrid and Brown, Peter",
title="A New Analysis of Fireball Data from the {M}eteorite {O}bservation and {R}ecovery {P}roject ({MORP})",
bookTitle="Modern Meteor Science An Interdisciplinary View",
year="2005",
publisher="Springer Netherlands",
address="Dordrecht",
pages="489--499",
isbn="978-1-4020-5075-6",
doi="10.1007/1-4020-5075-5_45",
url="https://doi.org/10.1007/1-4020-5075-5_45"
}

@article{RN664,
   author = {Moreno-Ibáñez, M. and Gritsevich, M. and Trigo-Rodríguez, J. M.},
   title = {New methodology to determine the terminal height of a fireball},
   journal = {Icarus},
   volume = {250},
   pages = {544-552},
   ISSN = {0019-1035},
   doi = {10.1016/j.icarus.2014.12.027},
   url = {<Go to ISI>://WOS:000349878200047},
   year = {2015},
   type = {Journal Article}
}

@article{PenaAsensio2023spanishnetwork,
    author = {Peña-Asensio, E and Trigo-Rodríguez, J M and Rimola, A and Corretgé-Gilart, M and Koschny, D},
    title = {Identifying meteorite droppers among the population of bright ‘sporadic’ bolides imaged by the {S}panish {M}eteor {N}etwork during the spring of 2022},
    journal = {Monthly Notices of the Royal Astronomical Society},
    volume = {520},
    number = {4},
    pages = {5173-5182},
    year = {2023},
    month = {01},
    issn = {0035-8711},
    doi = {10.1093/mnras/stad102},
    url = {https://doi.org/10.1093/mnras/stad102},
    eprint = {https://academic.oup.com/mnras/article-pdf/520/4/5173/49329818/stad102.pdf},
}

@INPROCEEDINGS{Millman1980spectra,
       author = {{Millman}, P.~M.},
        title = "{One hundred and fifteen years of meteor spectroscopy}",
    booktitle = {Solid Particles in the Solar System},
         year = 1980,
       editor = {{Halliday}, I. and {McIntosh}, B.~A.},
       series = {IAU Symposium},
       volume = {90},
        month = jan,
        pages = {121-127},
       adsurl = {https://ui.adsabs.harvard.edu/abs/1980IAUS...90..121M}
}

@article{Jenniskens2007spectra,
title = {Quantitative meteor spectroscopy: {E}lemental abundances},
journal = {Advances in Space Research},
volume = {39},
number = {4},
pages = {491-512},
year = {2007},
issn = {0273-1177},
doi = {https://doi.org/10.1016/j.asr.2007.03.040},
url = {https://www.sciencedirect.com/science/article/pii/S0273117707002633},
author = {P. Jenniskens}
}

@article{Halliday1996morp,
author = {Halliday, Ian and Griffin, Arthur A. and Blackwell, Alan T.},
title = {Detailed data for 259 fireballs from the {C}anadian camera network and inferences concerning the influx of large meteoroids},
journal = {Meteoritics \& Planetary Science},
volume = {31},
number = {2},
pages = {185-217},
doi = {https://doi.org/10.1111/j.1945-5100.1996.tb02014.x},
url = {https://onlinelibrary.wiley.com/doi/abs/10.1111/j.1945-5100.1996.tb02014.x},
eprint = {https://onlinelibrary.wiley.com/doi/pdf/10.1111/j.1945-5100.1996.tb02014.x},
year = {1996}
}

@article{Wesolowski2025cometanalog,
title = {Is it possible to create a realistic structure of a cometary nucleus analog in the laboratory?},
journal = {Icarus},
volume = {441},
pages = {116646},
year = {2025},
issn = {0019-1035},
doi = {https://doi.org/10.1016/j.icarus.2025.116646},
url = {https://www.sciencedirect.com/science/article/pii/S0019103525001939},
author = {Marcin Wesołowski and Zuzanna Bober and Łukasz Ożóg and Adrian Truszkiewicz and Maria Gritsevich and Mariusz Bester and Grzegorz Wisz}
}

@article{RN895,
   author = {Moreno-Ibáñez, M. and Gritsevich, M. and Trigo-Rodríguez, J. M. and Silber, E. A.},
   title = {Physically based alternative to the {PE} criterion for meteoroids},
   journal = {Monthly Notices of the Royal Astronomical Society},
   volume = {494},
   number = {1},
   pages = {316-324},
   ISSN = {0035-8711},
   doi = {10.1093/mnras/staa646},
   url = {<Go to ISI>://WOS:000535885900026},
   year = {2020},
   type = {Journal Article}
}

@article{OstrowskiBryson2019meteorites,
   author = {Ostrowski, D. and Bryson, K.},
   title = {The physical properties of meteorites},
   journal = {Planetary and Space Science},
   volume = {165},
   pages = {148-178},
   ISSN = {0032-0633},
   doi = {10.1016/j.pss.2018.11.003},
   url = {<Go to ISI>://WOS:000468702500014},
   year = {2019},
   type = {Journal Article}
}

@article{RN941,
   author = {Picone, J. M. and Hedin, A. E. and Drob, D. P. and Aikin, A. C.},
   title = {{NRLMSISE}-00 empirical model of the atmosphere: {S}tatistical comparisons and scientific issues},
   journal = {Journal of Geophysical Research-Space Physics},
   volume = {107},
   number = {A12},
   ISSN = {2169-9380},
   doi = {10.1029/2002ja009430},
   url = {<Go to ISI>://WOS:000181241900010},
   year = {2002},
   type = {Journal Article}
}

@article{RN782,
   author = {Popova, O. P. and Jenniskens, P. and Emel'yanenko, V. and Kartashova, A. and Biryukov, E. and Khaibrakhmanov, S. and Shuvalov, V. and Rybnov, Y. and Dudorov, A. and Grokhovsky, V. I. and Badyukov, D. D. and Yin, Q. Z. and Gural, P. S. and Albers, J. and Granvik, M. and Evers, L. G. and Kuiper, J. and Kharlamov, V. and Solovyov, A. and Rusakov, Y. S. and Korotkiy, S. and Serdyuk, I. and Korochantsev, A. V. and Larionov, M. Y. and Glazachev, D. and Mayer, A. E. and Gisler, G. and Gladkovsky, S. V. and Wimpenny, J. and Sanborn, M. E. and Yamakawa, A. and Verosub, K. L. and Rowland, D. J. and Roeske, S. and Botto, N. W. and Friedrich, J. M. and Zolensky, M. E. and Le, L. and Ross, D. and Ziegler, K. and Nakamura, T. and Ahn, I. and Lee, J. I. and Zhou, Q. and Li, X. H. and Li, Q. L. and Liu, Y. and Tang, G. Q. and Hiroi, T. and Sears, D. and Weinstein, I. A. and Vokhmintsev, A. S. and Ishchenko, A. V. and Schmitt-Kopplin, P. and Hertkorn, N. and Nagao, K. and Haba, M. K. and Komatsu, M. and Mikouchi, T. and Consortiu, Chelyabinsk Airburst},
   title = {Chelyabinsk Airburst, Damage Assessment, Meteorite Recovery, and Characterization},
   journal = {Science},
   volume = {342},
   number = {6162},
   pages = {1069-1073},
   ISSN = {0036-8075},
   doi = {10.1126/science.1242642},
   url = {<Go to ISI>://WOS:000327518600050},
   year = {2013},
   type = {Journal Article}
}

@article{RN198,
   author = {ReVelle, D. O. and Brown, P. G. and Spurn{\'y}, P.},
   title = {Entry dynamics and acoustics/infrasonic/seismic analysis for the {N}euschwanstein meteorite fall},
   journal = {Meteoritics and Planetary Science},
   volume = {39},
   number = {10},
   pages = {1605-1626},
   ISSN = {1086-9379},
   doi = {10.1111/j.1945-5100.2004.tb00061.x},
   url = {<Go to ISI>://WOS:000224852000002},
   year = {2004},
   type = {Journal Article}
}

@book{Jenniskens2023meteorshoweratlas,
  title={Atlas of Earth's meteor showers},
  author={Jenniskens, Peter},
  year={2023},
  publisher={Elsevier}
}

@article{Borovicka2026etavirginids,
    title = {Eta-{V}irginids: {A}nother asteroidal meteoroid stream},
    journal = {Icarus},
    volume = {450},
    pages = {116980},
    year = {2026},
    issn = {0019-1035},
    doi = {https://doi.org/10.1016/j.icarus.2026.116980},
    url = {https://www.sciencedirect.com/science/article/pii/S0019103526000461},
    author = {Jiří Borovička and Pavel Spurný and Pavel Koten and Gabriel {Borderes Motta} and Lenka Kotková and Rostislav Štork and Dušan Tomko and Thomas Weiland}
}

@article{Tancredi2014asteroidvscometorbits,
title = {A criterion to classify asteroids and comets based on the orbital parameters},
journal = {Icarus},
volume = {234},
pages = {66-80},
year = {2014},
issn = {0019-1035},
doi = {https://doi.org/10.1016/j.icarus.2014.02.013},
url = {https://www.sciencedirect.com/science/article/pii/S0019103514000992},
author = {Gonzalo Tancredi}
}

@article{RN892,
   author = {Rudawska, R. and Tóth, J. and Kalmancok, D. and Zigo, P. and Matlovič, P.},
   title = {Meteor spectra from {AMOS} video system},
   journal = {Planetary and Space Science},
   volume = {123},
   pages = {25-32},
   ISSN = {0032-0633},
   doi = {10.1016/j.pss.2015.11.018},
   url = {<Go to ISI>://WOS:000373414900004},
   year = {2016},
   type = {Journal Article}
}

@article{RN1108,
   author = {Sansom, E. K. and Bland, P. A. and Rutten, M. G. and Paxman, J. and Towner, M. C.},
   title = {Filtering Meteoroid Flights Using Multiple Unscented {K}alman {F}ilters},
   journal = {Astronomical Journal},
   volume = {152},
   number = {5},
   ISSN = {0004-6256},
   doi = {10.3847/0004-6256/152/5/148},
   url = {<Go to ISI>://WOS:000387557200002},
   year = {2016},
   type = {Journal Article}
}

@article{RN423,
   author = {Sansom, E. K. and Bland, P. A. and Towner, M. C. and Devillepoix, H. A. R. and Cupák, M. and Howie, R. M. and Jansen-Sturgeon, T. and Cox, M. A. and Hartig, B. A. D. and Paxman, J. P. and Benedix, G. and Forman, L.},
   title = {Murrili meteorite's fall and recovery from {K}ati {T}handa},
   journal = {Meteoritics and Planetary Science},
   volume = {55},
   number = {9},
   pages = {2157-2168},
   ISSN = {1086-9379},
   doi = {10.1111/maps.13566},
   url = {<Go to ISI>://WOS:000573878100001},
   year = {2020},
   type = {Journal Article}
}

@article{Moilanen2021strewnfields,
    author = {Moilanen, J. and Gritsevich, M. and Lyytinen, E.},
    title = {Determination of strewn fields for meteorite falls},
    journal = {Monthly Notices of the Royal Astronomical Society},
    volume = {503},
    number = {3},
    pages = {3337-3350},
    year = {2021},
    month = {03},
    issn = {0035-8711},
    doi = {10.1093/mnras/stab586},
    url = {https://doi.org/10.1093/mnras/stab586},
    eprint = {https://academic.oup.com/mnras/article-pdf/503/3/3337/36847146/stab586.pdf},
}

@article{Sansom2019alphabeta,
   author = {Sansom, E. K. and Gritsevich, M. and Devillepoix, H. A. R. and Jansen-Sturgeon, T. and Shober, P. and Bland, P. A. and Towner, M. C. and Cupák, M. and Howie, R. M. and Hartig, B. A. D.},
   title = {Determining Fireball Fates Using the alpha-beta Criterion},
   journal = {Astrophysical Journal},
   volume = {885},
   number = {2},
   ISSN = {0004-637x},
   doi = {10.3847/1538-4357/ab4516},
   url = {<Go to ISI>://WOS:000499886700001},
   year = {2019},
   type = {Journal Article}
}

@article{RN1224,
   author = {Shober, P. M.},
   title = {Solar heating and atmospheric filtering bias the meteorite record},
   journal = {Nature Astronomy},
   ISSN = {2397-3366},
   doi = {10.1038/s41550-025-02527-5},
   url = {<Go to ISI>://WOS:001466400100001},
   year = {2025},
   type = {Journal Article}
}

@article{RN400,
   author = {Shober, P. M. and Devillepoix, H. A. R. and Sansom, E. K. and Towner, M. C. and Cupák, M. and Anderson, S. L. and Benedix, G. and Forman, L. and Bland, P. A. and Howie, R. M. and Hartig, B. A. D. and Laubenstein, M. and Cary, F. and Langendam, A.},
   title = {Arpu {K}uilpu: {A}n {H}5 from the outer {M}ain {B}elt},
   journal = {Meteoritics and Planetary Science},
   volume = {57},
   number = {6},
   pages = {1146-1157},
   ISSN = {1086-9379},
   doi = {10.1111/maps.13813},
   url = {<Go to ISI>://WOS:000786537700001},
   year = {2022},
   type = {Journal Article}
}

@article{RN420,
   author = {Shober, P. M. and Jansen-Sturgeon, T. and Bland, P. A. and Devillepoix, H. A. R. and Sansom, E. K. and Towner, M. C. and Cupák, M. and Howie, R. M. and Hartig, B. A. D.},
   title = {Using atmospheric impact data to model meteoroid close encounters},
   journal = {Monthly Notices of the Royal Astronomical Society},
   volume = {498},
   number = {4},
   pages = {5240-5250},
   ISSN = {0035-8711},
   doi = {10.1093/mnras/staa2559},
   url = {<Go to ISI>://WOS:000587755500045},
   year = {2020},
   type = {Journal Article}
}

@article{RN413,
   author = {Shober, P. M. and Sansom, E. K. and Bland, P. A. and Devillepoix, H. A. R. and Towner, M. C. and Cupák, M. and Howie, R. M. and Hartig, B. A. D. and Anderson, S. L.},
   title = {The {M}ain {A}steroid {B}elt: {T}he Primary Source of Debris on Comet-like Orbits},
   journal = {Planetary Science Journal},
   volume = {2},
   number = {3},
   doi = {10.3847/PSJ/abde4b},
   url = {<Go to ISI>://WOS:000913045700001},
   year = {2021},
   type = {Journal Article}
}

@article{RN69,
   author = {Spurný, P. and Bland, P. A. and Shrbený, L. and Towner, M. C. and Borovička, J. and Bevan, A. W. R. and Vaughan, D.},
   title = {The {M}ason {G}ully Meteorite Fall in {SW} {A}ustralia: {F}ireball Trajectory and Orbit from Photographic Records},
   journal = {Meteoritics and Planetary Science},
   volume = {46},
   pages = {A220-A220},
   ISSN = {1086-9379},
   url = {<Go to ISI>://WOS:000293094700437},
   year = {2011},
   type = {Journal Article}
}

@article{RN1695,
   author = {Spurný, P. and Borovička, J. and Haloda, J. and Shrbený, L. and Heinlein, D.},
   title = {Two Very Precisely Instrumentally Documented Meteorite Falls: {Ž}d'ár nad {S}ázavou and {S}tubenberg: {P}rediction and Reality.},
   journal = {Meteoritics and Planetary Science},
   volume = {51},
   pages = {A591-A591},
   ISSN = {1086-9379},
   url = {<Go to ISI>://WOS:000388662400449},
   year = {2016},
   type = {Journal Article}
}

@article{RN137,
   author = {Spurný, P. and Borovička, J. and Kac, J. and Kalenda, P. and Atanackov, J. and Kladnik, G. and Heinlein, D. and Grau, T.},
   title = {Analysis of instrumental observations of the {J}esenice meteorite fall on {A}pril 9, 2009},
   journal = {Meteoritics and Planetary Science},
   volume = {45},
   number = {8},
   pages = {1392-1407},
   ISSN = {1086-9379},
   doi = {10.1111/j.1945-5100.2010.01121.x},
   url = {<Go to ISI>://WOS:000284427400011},
   year = {2010},
   type = {Journal Article}
}

@misc{Porubcan2009,
      title={The {T}aurid complex meteor showers and asteroids}, 
      author={V. Porubčan and L. Kornoš and I. P. Williams},
      year={2009},
      eprint={0905.1639},
      archivePrefix={arXiv},
      primaryClass={astro-ph.EP},
      url={https://arxiv.org/abs/0905.1639}, 
}

@article{RN1719,
   author = {Spurný, P. and Borovička, J. and Shrbený, L.},
   title = {The {Ž}d'ár nad {S}ázavou meteorite fall: {F}ireball trajectory, photometry, dynamics, fragmentation, orbit, and meteorite recovery},
   journal = {Meteoritics and Planetary Science},
   volume = {55},
   number = {2},
   pages = {376-401},
   ISSN = {1086-9379},
   doi = {10.1111/maps.13444},
   url = {<Go to ISI>://WOS:000511850800001},
   year = {2020},
   type = {Journal Article}
}

@article{RN132,
   author = {Spurný, P. and Haloda, J. and Borovička, J. and Shrbený, L. and Halodová, P.},
   title = {Reanalysis of the {B}enešov bolide and recovery of polymict breccia meteorites: {O}ld mystery solved after 20 years},
   journal = {Astronomy and Astrophysics},
   volume = {570},
   ISSN = {0004-6361},
   doi = {10.1051/0004-6361/201424308},
   url = {<Go to ISI>://WOS:000344158500098},
   year = {2014},
   type = {Journal Article}
}

@article{RN1339,
   author = {Spurný, P. and Oberst, J. and Heinlein, D.},
   title = {Photographic observations of {N}euschwanstein, a second meteorite from the orbit of the {P}říbram chondrite},
   journal = {Nature},
   volume = {423},
   number = {6936},
   pages = {151-153},
   ISSN = {0028-0836},
   doi = {10.1038/nature01592},
   url = {<Go to ISI>://WOS:000182699600040},
   year = {2003},
   type = {Journal Article}
}

@article{RN1371,
   author = {Tabetah, M. E. and Melosh, H. J.},
   title = {Air penetration enhances fragmentation of entering meteoroids},
   journal = {Meteoritics and Planetary Science},
   volume = {53},
   number = {3},
   pages = {493-504},
   ISSN = {1086-9379},
   doi = {10.1111/maps.13034},
   url = {<Go to ISI>://WOS:000426517500010},
   year = {2018},
   type = {Journal Article}
}

@article{RN131,
   author = {Tóth, J. and Svoren, J. and Borovička, J. and Spurný, P. and Igaz, A. and Kornos, L. and Veres, P. and Husárik, M. and Koza, J. and Kucera, A. and Zigo, P. and Gajdos, S. and Világi, J. and Capek, D. and Krisandová, Z. and Tomko, D. and Silha, J. and Schunová, E. and Bodnárová, M. and Búzová, D. and Krejcová, T.},
   title = {The {K}ošice meteorite fall: {R}ecovery and strewn field},
   journal = {Meteoritics and Planetary Science},
   volume = {50},
   number = {5},
   pages = {853-863},
   ISSN = {1086-9379},
   doi = {10.1111/maps.12447},
   url = {<Go to ISI>://WOS:000354258400002},
   year = {2015},
   type = {Journal Article}
}

@article{RN1244,
   author = {Towner, M. C. and Bland, P. A. and Spurný, P. and Benedix, G. K. and Dyl, K. and Greenwood, R. C. and Gibson, J. and Franchi, I. A. and Shrbený, L. and Bevan, A. W. R. and Vaughan, D.},
   title = {Mason {G}ully: {T}he Second Meteorite Recovered by the {D}esert {F}ireball {N}etwork},
   journal = {Meteoritics and Planetary Science},
   volume = {46},
   pages = {A238-A238},
   ISSN = {1086-9379},
   url = {<Go to ISI>://WOS:000293094700472},
   year = {2011},
   type = {Journal Article}
}

@article{Boro2022b,
	author = {Borovička, J. and Spurný, P. and Shrbený, L.},
	title = {Data on 824 fireballs observed by the digital cameras of the {E}uropean {F}ireball {N}etwork in 2017–2018: {II}. {A}nalysis of orbital and physical properties of centimeter-sized meteoroids},
	doi= {10.1051/0004-6361/202244197},
	url= {https://doi.org/10.1051/0004-6361/202244197},
	journal = {Astronomy and Astrophysics},
	year = 2022,
	volume = 667,
	pages = "A158",
}

@article{RN1706,
   author = {Trigo-Rodríguez, J. M. and Borovička, J. and Spurný, P. and Ortiz, J. L. and Docobo, J. A. and Castro-Tirado, A. J. and Llorca, J.},
   title = {The {V}illalbeto de la {P}eña meteorite fall: {II}. {D}etermination of atmospheric trajectory and orbit},
   journal = {Meteoritics and Planetary Science},
   volume = {41},
   number = {4},
   pages = {505-517},
   ISSN = {1086-9379},
   doi = {10.1111/j.1945-5100.2006.tb00478.x},
   url = {<Go to ISI>://WOS:000237212300002},
   year = {2006},
   type = {Journal Article}
}

@article{RN286,
   author = {Vida, D. and Šegon, D. and Gural, P. S. and Brown, P. G. and McIntyre, M. J. M. and Dijkema, T. J. and Pavletic, L. and Kukic, P. and Mazur, M. J. and Eschman, P. and Roggemans, P. and Merlak, A. and Zubovic, D.},
   title = {The {G}lobal {M}eteor {N}etwork: {M}ethodology and first results},
   journal = {Monthly Notices of the Royal Astronomical Society},
   volume = {506},
   number = {4},
   pages = {5046-5074},
   ISSN = {0035-8711},
   doi = {10.1093/mnras/stab2008},
   url = {<Go to ISI>://WOS:000705339100027},
   year = {2021},
   type = {Journal Article}
}

@article{RN1304,
   author = {Vojácek, V. and Borovička, J. and Koten, P. and Spurný, P. and Štork, R.},
   title = {Catalogue of representative meteor spectra},
   journal = {Astronomy and Astrophysics},
   volume = {580},
   ISSN = {0004-6361},
   doi = {10.1051/0004-6361/201425047},
   url = {<Go to ISI>://WOS:000360020200067},
   year = {2015},
   type = {Journal Article}
}

@article{RN99,
   author = {Vojácek, V. and Borovička, J. and Koten, P. and Spurný, P. and Štork, R.},
   title = {Properties of small meteoroids studied by meteor video observations},
   journal = {Astronomy and Astrophysics},
   volume = {621},
   ISSN = {1432-0746},
   doi = {10.1051/0004-6361/201833289},
   url = {<Go to ISI>://WOS:000455270200002},
   year = {2019},
   type = {Journal Article}
}

@article{RN374,
   author = {Zappatini, A. and Hofmann, B. A. and Gnos, E. and Eggenberger, U. and Gfeller, F. and Kruttasch, P. M. and Sansom, E. K. and Devillepoix, H. A. R. and Cupák, M. and Lindemann, S. and Booz, B. and Al-Muati, M. S. and Al-Ghafri, H. A. and Al-Zakwani, A. A.},
   title = {Al-{K}hadhaf: {A} Camera-Observed {H}5-6 Fall from {O}man.},
   journal = {Meteoritics and Planetary Science},
   volume = {59},
   pages = {A457-A457},
   ISSN = {1086-9379},
   url = {<Go to ISI>://WOS:001317679600458},
   year = {2024},
   type = {Journal Article}
}

@article{RN1674,
   author = {Zinn, J. and Judd, O. D. P. and ReVelle, D. O.},
   title = {Leonid meteor ablation, energy exchange, and trail morphology},
   journal = {Impact of Minor Bodies of Our Solar System on Planets and Their Middle and Upper Atmosphere},
   volume = {33},
   number = {9},
   pages = {1466-1474},
   ISSN = {0273-1177},
   doi = {10.1016/j.asr.2003.04.001},
   url = {<Go to ISI>://WOS:000222364500004},
   year = {2004},
   type = {Journal Article}
}

@article{RN1558,
   author = {Zuluaga, J. I. and Cuartas-Restrepo, P. A. and Ospina, J. and Sucerquia, M.},
   title = {Can we predict the impact conditions of metre-sized meteoroids?},
   journal = {Monthly Notices of the Royal Astronomical Society},
   volume = {486},
   number = {1},
   pages = {L69-L73},
   ISSN = {0035-8711},
   doi = {10.1093/mnrasl/slz060},
   url = {<Go to ISI>://WOS:000482696100017},
   year = {2019},
   type = {Journal Article}
}

@article{egal2025catastrophic,
  title={Catastrophic disruption of asteroid 2023 {CX}1 and implications for planetary defence},
  author={Egal, Auriane and Vida, Denis and Colas, Fran{\c{c}}ois and Zanda, Brigitte and Bouley, Sylvain and Steinhausser, Asma and Vernazza, Pierre and Ferri{\`e}re, Ludovic and Gattacceca, J{\'e}r{\^o}me and Birlan, Mirel and others},
  journal={Nature Astronomy},
  pages={1--14},
  year={2025},
  publisher={Nature Publishing Group UK London}
}

@article{devillepoix2022meteor,
  title={A Meteor Spectroscopic Survey in the Nullarbor},
  author={Devillepoix, Hadrien AR and T{\'o}th, Juraj and Matlovi{\v{c}}, Pavol and Cup{\'a}k, Martin and Towner, Martin C and Sansom, Eleanor K and Korno{\v{s}}, Leonard and Paulech, Tom{\'a}{\v{s}} and Zigo, Pavol},
  journal={Research Notes of the AAS},
  volume={6},
  number={7},
  pages={144},
  year={2022},
  publisher={The American Astronomical Society}
}

\appendix
\numberwithin{equation}{section}
\numberwithin{table}{section}
\numberwithin{figure}{section}

\section{Derivation of Novel Formulae for Dynamical Analysis of MDOs} \label{sec:deriv}

The formulae we have used to estimate $\alpha$ and $\beta$ for `minimally decelerating' meteoroids are provided in Section \ref{subsec:neweqs}. Their algebraic derivation follows from formulae presented in prior independent studies. First, we employ the terminal altitude equation of \cite{RN696}, which implicitly assumes the absence of a terminal mass:

\begin{equation}
y_t = \ln(2 \alpha \beta)
\end{equation}

First, the ballistic coefficient $\alpha$ is found by substituting the defining equation for $\beta$ as follows:

\begin{equation}
\beta = \frac{(1 - \mu) \sigma_b V_e^2}{2}
\end{equation}

\begin{equation}
y_t = \ln(\alpha (1 - \mu) \sigma_b V_e^2)
\end{equation}

\begin{equation}
e^{y_t} = \alpha (1 - \mu) \sigma_b V_e^2
\end{equation}

Adopting the common assumption of uniform ablation across the meteoroid's surface, we set \(\mu = \frac{2}{3}\). The term (1-\(\mu\)) therefore becomes \(\frac{1}{3}\), leading to:

\begin{equation}
e^{y_t} = \frac{\alpha \sigma_b V_e^2}{3}
\end{equation}

\begin{equation}
\alpha = \frac{3e^{y_t}}{\sigma_b V_e^2}
\end{equation}

Once $\alpha$ is known, the mass loss parameter $\beta$ can be found by rearranging the terminal altitude formula (B6):

\begin{equation}
\beta = \frac{e^{y_t}}{2 \alpha}
\end{equation}

Into this we may substitute the novel formula for $\alpha$, to arrive at a novel formula for $\beta$ that does not depend on the normalised terminal altitude:

\begin{equation}
\beta = \frac{e^{y_t} \sigma_b V_e^2}{6e^{y_t}}
\end{equation}

\begin{equation}
\beta = \frac{\sigma_b V_e^2}{6}
\end{equation}

In this scheme, the values of both $\alpha$ and $\beta$ depend upon the bulk ablation coefficient \(\sigma_b\). It is for this reason that we have devoted Section \ref{subsec:sigmameth} to placing reasonable constraints on ablation coefficients for metallic, stony and cometary meteoroid types.

\section{Data Tables}\label{sec:data}

\begin{sidewaystable*}[ht]\label{tab:meteorites1}
    \small
    \begin{center}
    \caption{Kinematic and dynamic parameters for a selection of instrumentally observed meteorite falls. * indicates a provisional name. $\alpha$: Ballistic coefficient; $\beta$: Mass loss parameter; $\gamma$: Entry angle measured from horizontal; $V_e$: Entry velocity; $H_t$: Terminal altitude; $\sigma_b$: Bulk ablation coefficient. Table continues on following page.}
    \begin{tabular}{|l|l|r|r|r|r|r|r|r|l|l|}
    \hline
     Meteorite & Type & $\alpha$ & $\beta$ & $\gamma$ (°) & \(V_e\) (km s\(^{-1}\)) & \(H_t\) (km) & ln(\(\alpha\)sin\(\gamma\)) & ln(\(\beta\)) & \(\sigma_b\) (kg MJ\(^{-1}\)) & Reference(s) \\
    \hline
     Ådalen* & Iron & 0.88 & 0.50 & 70.4 & 17.4 & 11.3 & -0.19 & -0.69 & 0.0099 & \cite{RN543} \\
     \null & \null & \null & \null & \null & \null & \null & \null & \null & \null & \cite{Moilanen2026adalen} \\
    \hline
     Al-Khadhaf & H5/6 & 26.06 & 0.66 & 68.5 & 14.0 & 30.2 & 3.19 & -0.42 & 0.0200 & \cite{RN374} \\
     \null & \null & \null & \null & \null & \null & \null & \null & \null & \null & \cite{Zappatini2026alkhadhaf} \\
    \hline
     Arpu Kuilpu & H5 & 40.84 & 0.95 & 49.7 & 17.9 & 28.9 & 3.44 & -0.05 & 0.0179 & \cite{RN400} \\
     \null & \null & \null & \null & \null & \null & \null & \null & \null & \null & \cite{RN371} \\
    \hline
     Benešov & LL3/4 & 7.15 & 1.77 & 80.7 & 21.3 & 19.5 & 1.95 & 0.57 & 0.0235 & \cite{RN1354} \\
     \null & \null & \null & \null & \null & \null & \null & \null & \null & \null & \cite{Gritsevich2008fireballdynamics} \\
     \null & \null & \null & \null & \null & \null & \null & \null & \null & \null & \cite{RN132} \\
    \hline
     Cavezzo & L5 & 18.75 & 0.30 & 68.4 & 12.2 & 21.5 & 2.86 & -1.21 & 0.0120 & \cite{Gardiol2021cavezzo} \\
    \hline
     Chelyabinsk & LL5 & 0.35 & 2.25 & 18.3 & 19.2 & 12.6 & -2.21 & 0.81 & 0.0368 & \cite{Trigo2021} \\
     \null & \null & \null & \null & \null & \null & \null & \null & \null & \null & \cite{RN782} \\
     \null & \null & \null & \null & \null & \null & \null & \null & \null & \null & \cite{RN1558} \\
    \hline
     Dingle Dell & LL6 & 13.87 & 0.62 & 51.4 & 15.4 & 19.1 & 2.38 & -0.49 & 0.0155 & \cite{RN436} \\
     \null & \null & \null & \null & \null & \null & \null & \null & \null & \null & \cite{RN432} \\
    \hline
     Golden & L5/LL5 & 16.07 & 0.61 & 54.3 & 17.9 & 18.5 & 2.57 & -0.50 & 0.0114 & \cite{RN162} \\
    \hline
     Grimsby & H5 & 15.50 & 0.90 & 55.2 & 20.9 & 19.6 & 2.54 & -0.11 & 0.0124 & \cite{RN158} \\
    \hline
     Hamburg & H4 & 7.93 & 1.87 & 66.0 & 15.8 & 19.7 & 1.98 & 0.63 & 0.0448 & \cite{RN164} \\
    \hline
     Ischgl & LL6 & 20.29 & 1.62 & 19.4 & 21.3 & 38.1 & 1.91 & 0.48 & 0.0214 & \cite{RN538} \\
    \hline
     Innisfree & L5 & 7.66 & 1.92 & 67.8 & 14.5 & 20.0 & 1.96 & 0.65 & 0.0548 & \cite{REVELLE1979453} \\
     \null & \null & \null & \null & \null & \null & \null & \null & \null & \null & \cite{Gritsevich2008} \\
    \hline
     Jesenice & L6 & 8.24 & 0.82 & 58.8 & 13.7 & 18.0 & 1.95 & -0.20 & 0.0261 & \cite{RN137} \\
    \hline
     Košice & H5 & 3.21 & 3.15 & 59.8 & 15.0 & 17.4 & 1.02 & 1.15 & 0.0846 & \cite{Kuznetsova2014kosicetrajectory} \\
     \null & \null & \null & \null & \null & \null & \null & \null & \null & \null & \cite{RN131} \\
     \null & \null & \null & \null & \null & \null & \null & \null & \null & \null & \cite{Moilanen2021strewnfields} \\
    \hline
     Križevci & H6 & 13.62 & 1.18 & 65.4 & 18.2 & 21.9 & 2.52 & 0.17 & 0.0214 & \cite{Boro2015krizevci} \\
    \hline
     Kybo-Lintos* & H4/5 & 19.87 & 1.33 & 63.9 & 25.3 & 25.2 & 2.88 & 0.28 & 0.0124 & \cite{RN3} \\
    \hline
     La Posa Plain & LL3-6 & 14.83 & 0.74 & 49.8 & 13.5 & 30.6 & 2.43 & -0.31 & 0.0241 & \cite{RN1753} \\
    \hline
     Lost City & H5 & 11.34 & 1.13 & 35.0 & 14.2 & 19.0 & 1.87 & 0.12 & 0.0338 & \cite{REVELLE1979453} \\
     \null & \null & \null & \null & \null & \null & \null & \null & \null & \null & \cite{Gritsevich2008} \\
    \hline
    \end{tabular}
    \end{center}
\end{sidewaystable*}

\begin{sidewaystable*}\label{tab:meteorites2}
    \small
    \begin{center}
    \begin{tabular}{|l|l|r|r|r|r|r|r|r|l|l|}
    \hline
     Meteorite & Type & $\alpha$ & $\beta$ & $\gamma$ (°) & \(V_e\) (km s\(^{-1}\)) & \(H_t\) (km) & ln(\(\alpha\)sin\(\gamma\)) & ln(\(\beta\)) & \(\sigma_b\) (kg MJ\(^{-1}\)) & Reference(s) \\
    \hline
     Madura Cave & L5 & 10.87 & 0.68 & 58.3 & 14.3 & 18.6 & 2.26 & -0.39 & 0.0198 & \cite{RN1229} \\
     \null & \null & \null & \null & \null & \null & \null & \null & \null & \null & \cite{RN5} \\
    \hline
     Maribo & CM2 & 8.20 & 5.74 & 31.7 & 28.3 & 30.6 & 1.46 & 1.75 & 0.0430 & \cite{RN127} \\
    \hline
     Mason Gully & H5 & 9.08 & 0.95 & 53.9 & 14.5 & 23.8 & 1.99 & -0.06 & 0.0269 & \cite{RN69} \\
     \null & \null & \null & \null & \null & \null & \null & \null & \null & \null & \cite{RN1244} \\
    \hline
     Morávka & H5 & 15.50 & 1.16 & 20.4 & 21.9 & 21.0 & 1.69 & 0.15 & 0.0145 & \cite{Boro2003moravkatrajectory} \\
     \null & \null & \null & \null & \null & \null & \null & \null & \null & \null & \cite{BoroKalenda2003moravkafragmentation} \\
    \hline
     Murrili & H5 & 5.50 & 1.06 & 68.5 & 13.7 & 18.3 & 1.63 & 0.06 & 0.0340 & \cite{RN423} \\
    \hline
     Neuschwanstein & EL6 & 3.92 & 2.59 & 49.5 & 21.0 & 16.0 & 1.09 & 0.95 & 0.0353 & \cite{RN198} \\
     \null & \null & \null & \null & \null & \null & \null & \null & \null & \null & \cite{RN1339} \\
     \null & \null & \null & \null & \null & \null & \null & \null & \null & \null & \cite{RN694} \\
    \hline
     Park Forest & L5 & 3.31 & 2.23 & 61.0 & 19.5 & 18.0 & 1.06 & 0.80 & 0.0351 & \cite{RN1745} \\
    \hline
     Puli Ilkaringuru & H5 & 19.90 & 1.35 & 32.5 & 17.9 & 25.1 & 2.37 & 0.30 & 0.0252 & \cite{RN1753} \\
    \hline
     Raja & EH3 & 23.87 & 0.81 & 59.1 & 12.8 & 25.9 & 3.02 & -0.21 & 0.0296 & \cite{RN1753} \\
    \hline
     Saint-Pierre-le-Viger & L5/6 & 15.98 & 0.49 & 48.6 & 14.0 & 18.9 & 2.48 & -0.71 & 0.0150 & \cite{RN1691} \\
     \null & \null & \null & \null & \null & \null & \null & \null & \null & \null & \cite{egal2025catastrophic} \\
    \hline
     Stubenberg & LL6 & 9.08 & 0.95 & 70.0 & 13.9 & 17.6 & 2.14 & -0.05 & 0.0295 & \cite{RN1695} \\
     \null & \null & \null & \null & \null & \null & \null & \null & \null & \null & \cite{RN1694} \\
    \hline
     Taghzout & H5 & 18.84 & 1.16 & 29.3 & 17.1 & 25.2 & 2.22 & 0.15 & 0.0239 & \cite{RN1696} \\
     \null & \null & \null & \null & \null & \null & \null & \null & \null & \null & \cite{RN1697} \\
    \hline
     Tagish Lake & C2-ung & 14.00 & 2.25 & 17.8 & 15.8 & 29.0 & 1.45 & 0.81 & 0.0540 & \cite{RN1702} \\
     \null & \null & \null & \null & \null & \null & \null & \null & \null & \null & \cite{RN1700} \\
     \null & \null & \null & \null & \null & \null & \null & \null & \null & \null & \cite{Ceplecha2007tagishlakefragmentation} \\
    \hline
     Takapō & L5 & 16.49 & 1.51 & 44.6 & 18.8 & 25.0 & 2.45 & -0.67 & 0.0257 & \cite{RN1753} \\
    \hline
     Traspena & L5 & 3.19 & 2.35 & 76.7 & 15.1 & 15.8 & 1.13 & 0.85 & 0.0620 & \cite{RN1705} \\
    \hline
     Villalbeto de la Peña & L6 & 6.28 & 2.36 & 29.0 & 16.9 & 22.0 & 1.11 & 0.86 & 0.0496 & \cite{RN1706} \\
    \hline
     Winchcombe & CM2 & 37.51 & 0.86 & 41.8 & 13.9 & 27.6 & 3.22 & -0.12 & 0.0276 & \cite{RN1716} \\
     \null & \null & \null & \null & \null & \null & \null & \null & \null & \null & \cite{RN1711} \\
    \hline
     Žd'ár nad Sázavou & L4 & 21.06 & 1.35 & 25.7 & 21.9 & 24.7 & 2.21 & 0.30 & 0.0169 & \cite{RN1695} \\
     \null & \null & \null & \null & \null & \null & \null & \null & \null & \null & \cite{RN1719} \\
    \hline
    \end{tabular}
    \end{center}
\end{sidewaystable*}

\begin{table*}\label{tab:irons}
    \begin{center}
    \caption{Kinematic and dynamic parameters for a selection of instrumentally observed `minimally decelerating' iron meteoroids whose emission spectra have been captured by AMOS in conjunction with the GFO or EFN \citep{RN1304, RN99, RN890, Boro2022b}. In all cases, an ablation coefficient of 0.074 kg MJ\(^{-1}\) has been assumed \citep{ReVelleCeplecha1994}. Fireball DN240406\(\_\)02 has not previously been reported.}
    \begin{tabular}{|l|r|r|r|r|r|r|r|}
    \hline
    Designation & $\alpha$ & $\beta$ & \(\gamma\) (°) & \(V_e\) (km s\(^{-1}\)) & \(H_t\) (km) & ln(\(\alpha\)sin\(\gamma\)) & ln(\(\beta\)) \\
    \hline
    DN 240406\(\_\)02 & 129 & 12.13 & 63.0 & 13.1 & 56.3 & 4.73 & 2.50 \\
    EN 260217\(\_\)193948 & 423 & 1.85 & 37.1 & 12.2 & 52.7 & 5.54 & 0.62 \\
    EN 180417\(\_\)000218 & 1814 & 7.76 & 29.1 & 25.0 & 73.4 & 6.78 & 2.05 \\
    EN 200917\(\_\)232223 & 632 & 3.31 & 45.6 & 16.3 & 59.7 & 6.12 & 1.20 \\
    EN 020618\(\_\)001354 & 509 & 14.48 & 29.2 & 34.2 & 68.7 & 5.51 & 2.67 \\
    AMOS 08505025 & 5999 & 3.21 & 60.0 & 16.1 & 75.6 & 8.56 & 1.17 \\
    AMOS 06406048 & 3157 & 21.74 & 32.3 & 41.9 & 84.7 & 7.43 & 3.08 \\
    AMOS 06C14187 & 3999 & 2.00 & 50.5 & 12.7 & 69.3 & 8.04 & 0.69 \\
    AMOS 08505008 & 3533 & 8.76 & 58.9 & 26.6 & 79.0 & 8.02 & 2.17 \\
    AMOS 10406060 & 2860 & 13.16 & 33.0 & 32.6 & 80.4 & 7.35 & 2.58 \\
    \hline
    \end{tabular}
    \end{center}
\end{table*}

\end{document}